\documentclass[twocolumn,showpacs,preprintnumbers,amsmath,amssymb]{revtex4}

\usepackage{graphicx}
\usepackage{dcolumn}
\usepackage{bm}

\begin{document}

\title{Nucleon-induced reactions at intermediate energies: New data at 96 MeV and theoretical status}

\author{V. Blideanu$^{1}$\footnote{blideanu@lpccaen.in2p3.fr}, 
        F. R. Lecolley$^{1}$, J. F. Lecolley$^{1}$, T. Lefort$^{1}$, N. Marie$^{1}$,
	A. Ata\c{c}$^{2}$, G. Ban$^{1}$,\\
	B. Bergenwall$^{2}$, J. Blomgren$^{2}$, S. Dangtip$^{2,3}$,
	K. Elmgren$^{4}$, Ph. Eudes$^{5}$, Y. Foucher$^{6}$, A. Guertin$^{5}$,\\
	F. Haddad$^{5}$,
	A. Hildebrand$^{2}$, C. Johansson$^{2}$, O. Jonsson$^{7}$, M. Kerveno$^{8}$, T. Kirchner$^{5}$,
	J. Klug$^{2}$,\\
	Ch. Le Brun$^{9}$, C. Lebrun$^{5}$, M. Louvel$^{1}$, P. Nadel-Turonski$^{10}$,
	L. Nilsson$^{2,7}$, N. Olsson$^{2,4}$, S. Pomp$^{2}$,\\
	A. V. Prokofiev$^{7}$, P.-U. Renberg$^{7}$,
	G. Rivi\`ere$^{5}$, I. Slypen$^{11}$, L. Stuttg\'e$^{8}$, U. Tippawan$^{2,3}$, M. \"Osterlund$^{2}$}
				     
\affiliation{$^{1}$LPC, ENSICAEN, Universit\'e de Caen, CNRS/IN2P3, Caen, France}
\affiliation{$^{2}$Department of Neutron Research, Uppsala University, Sweden}
\affiliation{$^{3}$Fast Neutron Research Facility, Chiang Mai University, Thailand}
\affiliation{$^{4}$Swedish Defence Research Agency, Stockholm, Sweden}
\affiliation{$^{5}$SUBATECH, Universit\'e de Nantes, CNRS/IN2P3, France}
\affiliation{$^{6}$DSM/DAPNIA/SPhN, CEA Saclay, Gif-sur-Yvette, France}
\affiliation{$^{7}$The Svedberg Laboratory, Uppsala University, Sweden}
\affiliation{$^{8}$IReS, Strasbourg, France}
\affiliation{$^{9}$Laboratoire de Physique Subatomique et de Cosmologie, Grenoble, France}
\affiliation{$^{10}$Department of Radiation Sciences, Uppsala University, Sweden}
\affiliation{$^{11}$Institut de Physique Nucl\'eaire, Universit\'e Catholique de Louvain, Louvain-la-Neuve, Belgium}      

\begin{abstract}
Double-differential cross sections for light charged particle production (up to $A=4$) were measured 
in 96 MeV neutron-induced reactions, at TSL laboratory cyclotron in Uppsala (Sweden). Measurements
for three targets, Fe, Pb, and U, were performed using two independent devices, SCANDAL and MEDLEY.
The data were recorded with low energy thresholds and for a wide angular range ($20-160$ degrees).
The normalization procedure used to extract the cross sections 
is based on the \textit{np} elastic scattering reaction that we
measured and for which we present experimental results. A good control of the systematic 
uncertainties affecting the results is achieved. Calculations using the exciton model are reported. 
Two different theoretical approches proposed to improve its predictive power regarding the complex
particle emission are tested. The capabilities of each approach is illustrated by
comparison with the 96 MeV data that we measured, and with other experimental results available
in the literature.
\end{abstract}

	
\maketitle				  

\section{Introduction}
The deep understanding of nucleon-induced reactions is a crucial step for 
the further development of nuclear reactions theory in general. In addition, a
complete information in this field is strongly needed for a large amount of applications,
such as the incineration of nuclear waste with accelerator-driven 
systems (ADS), cancer therapy or the control of radiation effects induced 
by terrestrial cosmic rays in microelectronics. For this reason, 
the problem of nucleon-induced reactions has gained a renewed 
interest in the last few years. This interest has been expressed in part 
by extensive experimental campaigns, such as, for example, those carried out by
several laboratories in Europe in the framework of the HINDAS program \cite{Hi00}.

Particularly, nucleon-induced reactions in the 20-200 MeV energy range, 
have for a long time been the subject of intensive theoretical studies.
For this range, the first major step for the improvement of nuclear reaction models
consisted of the introduction of the so-called "pre-equilibrium process". This process
has been proposed in order to explain the smooth dependence of the particle emission 
probability versus angle and energy, which has been observed experimentally.
This "pre-equilibrium" process is supposed to occur at an intermediate stage and to consist
of multiple nucleon-nucleon interactions which take place inside the target 
nucleus. During that process, particle emission occurs 
after completion of the one-step interaction phase, i.e., direct process phase, 
but a long time before the statistical equilibrium of the compound nucleus has been reached.

\begin{figure*}
\begin{center}
\includegraphics[width=0.6\textwidth]{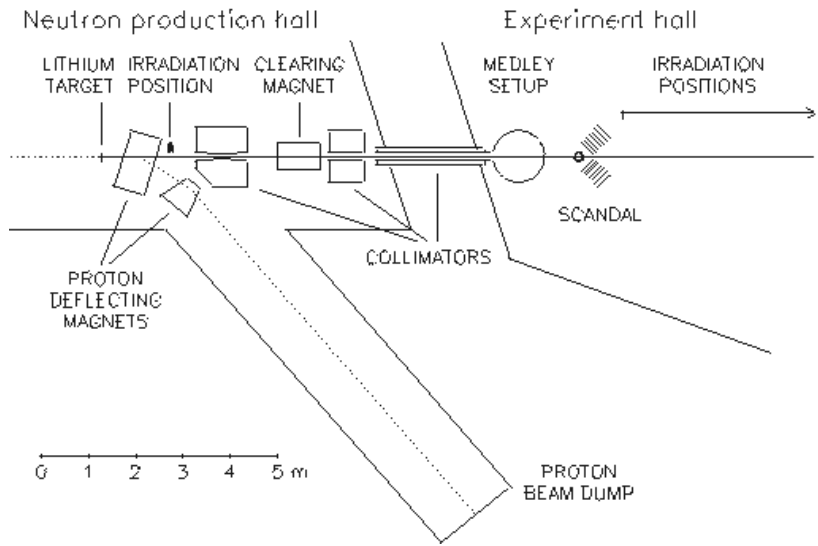}
\caption{\label{fig:epsart}TSL neutron beam facility and the location of
the detection systems used in the experiment.}
\end{center}
\end{figure*}

During the last 40 years, several approaches attempted to give a 
theoretical description of this pre-equilibrium process. 
Some of them have shown all along a good predictive power 
for a wide set of experimental energy distributions of nucleons emitted
in nucleon-nucleus reactions. However, those models were unable to reproduce the experimental 
distributions of complex particles, for which they systematically underestimate
the production rates. Among them, the exciton model of Griffin \cite{Gr66} is a very good example.
Originally introduced in 1966, this model 
has been quickly adopted by the community because of
its adaptability and simplicity. In an attempt to increase its capability in reproducing 
the complex particle rates, two main approaches have been developed.
The first one, proposed in 1973 \cite{Ri73}, introduces
a cluster formation probability during the nucleon-nucleon interactions inside the nucleus.
The second one formulated by Kalbach in 1977 \cite{Kal77}  is a 
completely different approach which takes into account the possible 
contributions of direct pick-up and knock-out mechanisms. 

Nowadays, the exciton model modified according to these theories is 
the only one available to calculate energy spectra of both nucleons and 
complex particles emitted in nucleon-induced reactions at intermediate
energies. In the past, both approaches have been tested against data, 
and they both show a satisfactory agreement with experimental
distributions \cite{Kal77,Wu78}. The comparisons 
were made using the data available at that moment and which concern 
a limited number of reaction configurations and incident energies, lower 
than 63 MeV. Despite this success, several questions are still open to 
discussion. An extended study of the influence of the entrance channel parameters is necessary, i.e,
the dependence on the incident particle type and on the incident energy has to be investigated.

The measurements presented in this work are part of the HINDAS program and
they concern double-differential distributions of light charged particles, up to $A=4$, 
emitted in 96 MeV neutron induced reactions on three targets, iron, lead and uranium. 
Calculations for those reactions are performed with the basic exciton model
implemented in the GNASH code \cite{Ch92}, and with both independent approaches proposed 
respectively by Ribansk\'y-Oblo\v zinsk\'y~\cite{Ri73} 
and by Kalbach~\cite{Kal77}. The robustness of those approaches are also tested
for other reactions with incident protons at lower energies and with other targets
for which experimental results are available in the literature. 
This study allows a global view on the predictive power of each model.

The experimental set-up used for the data taking is briefly presented 
in section II. In section III, details concerning the
procedures used to obtain the energy spectra 
and the cross section normalization are given.  The results are 
presented in section IV. Section V is dedicated to the description 
of the theoretical calculations related to the particle emission in nucleon-induced
reactions at intermediate energies, and the predictions of the models are compared to 
experimental data. The conclusions of this work are given in section VI.

\section{Experimental procedure}
The experiments were performed using the neutron beam available at the TSL laboratory
in Uppsala (Sweden) whose facility is presented in Fig. 1. 
Neutrons were produced by $^{7}$Li(p,n)$^{7}$Be reactions using a 100 MeV proton 
beam impinging on a lithium target. The beam monitoring was provided 
by a Faraday cup where the proton beam was dumped and by a fission detector
composed of thin-film breakdown counters~\cite{Smi95} placed in the experimental hall. 
The stability of 
the beam was checked regularly during the data taking. The deviations found
between the indications of both monitors did not exceeded 2 \%. 

Difficulties encountered when working with neutron beams are related to
their characteristics. The neutrons of the beam are not strictly
mono-energetic. This is illustrated in Fig. 2(a) where a typical neutron spectrum is shown. 
It presents two components: one is a peak centered at an energy slighty lower than the 
incident proton beam energy ($Q=-1.6 MeV$), diminished of the energy loss inside the production 
target, 
and the other is a low energy tail which contains about 50 \% of the total 
number of produced neutrons, and which originates from highly excited
states of $^{7}$Be. For the data analysis,
events associated with low energy neutrons must be rejected. The method employed for this
rejection will be described in the forthcoming sections. After selection, the intensity of 
the resulting 96 MeV neutron beam is of the order of 10$^4$ n/cm$^2$/sec. The neutron 
beam is collimated to a solid angle of 60~$\mu$sr and the beam spot at about 10 meters 
from the lithium target has a diameter of 8 cm (Fig. 2(b)). These characteristics
impose the use of an adequate detection set-up in order to obtain a 
satisfactory counting rate, keeping, at the same time, the energy and angular 
resolutions within reasonable limits. 
Two independent detection systems, MEDLEY and SCANDAL, were used in our
experiments. They were placed one after the other on the beam line as shown in Fig. 1.

\begin{figure}
\begin{center}
\includegraphics[width=0.23\textwidth]{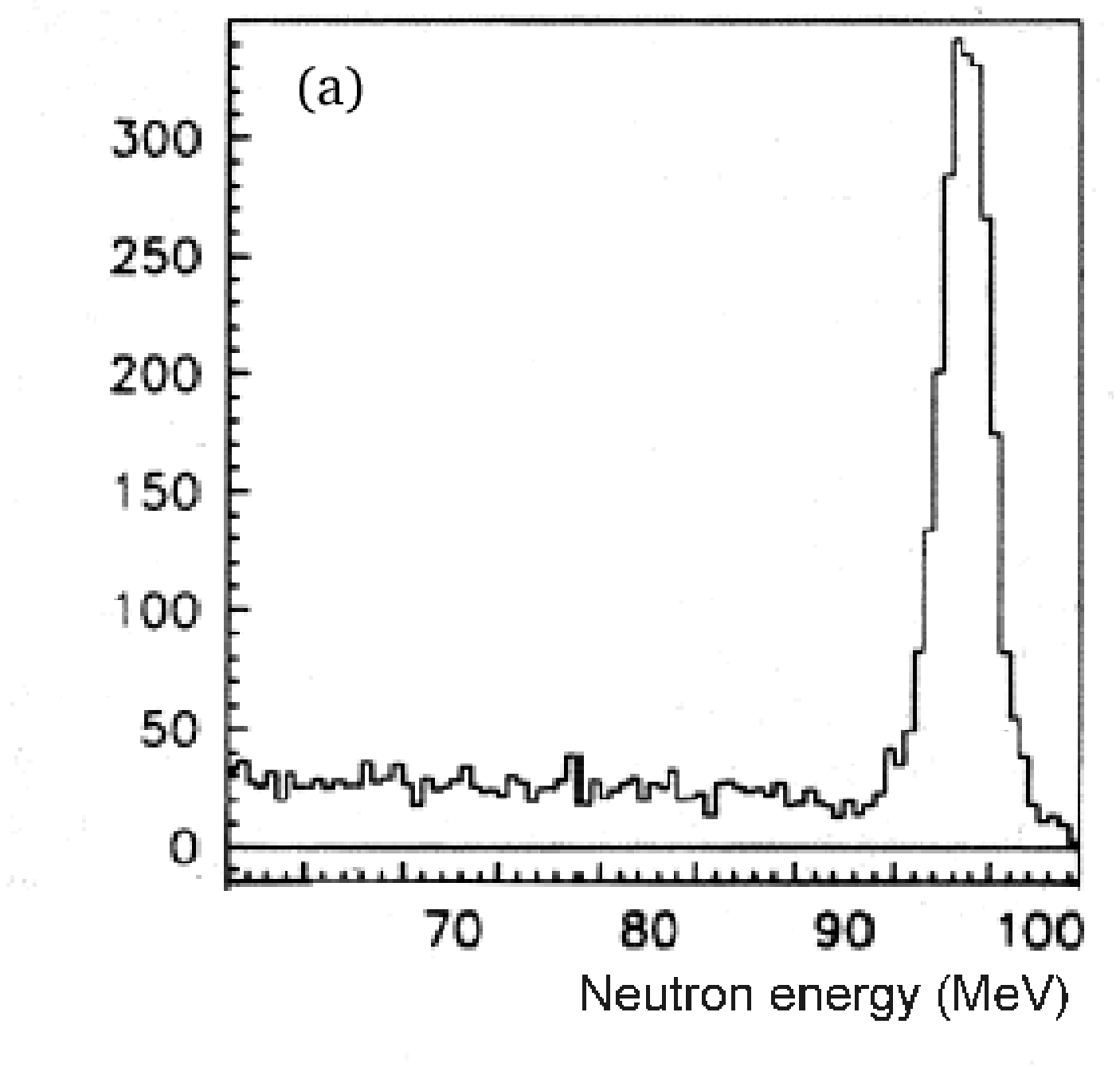}
\includegraphics[width=0.23\textwidth]{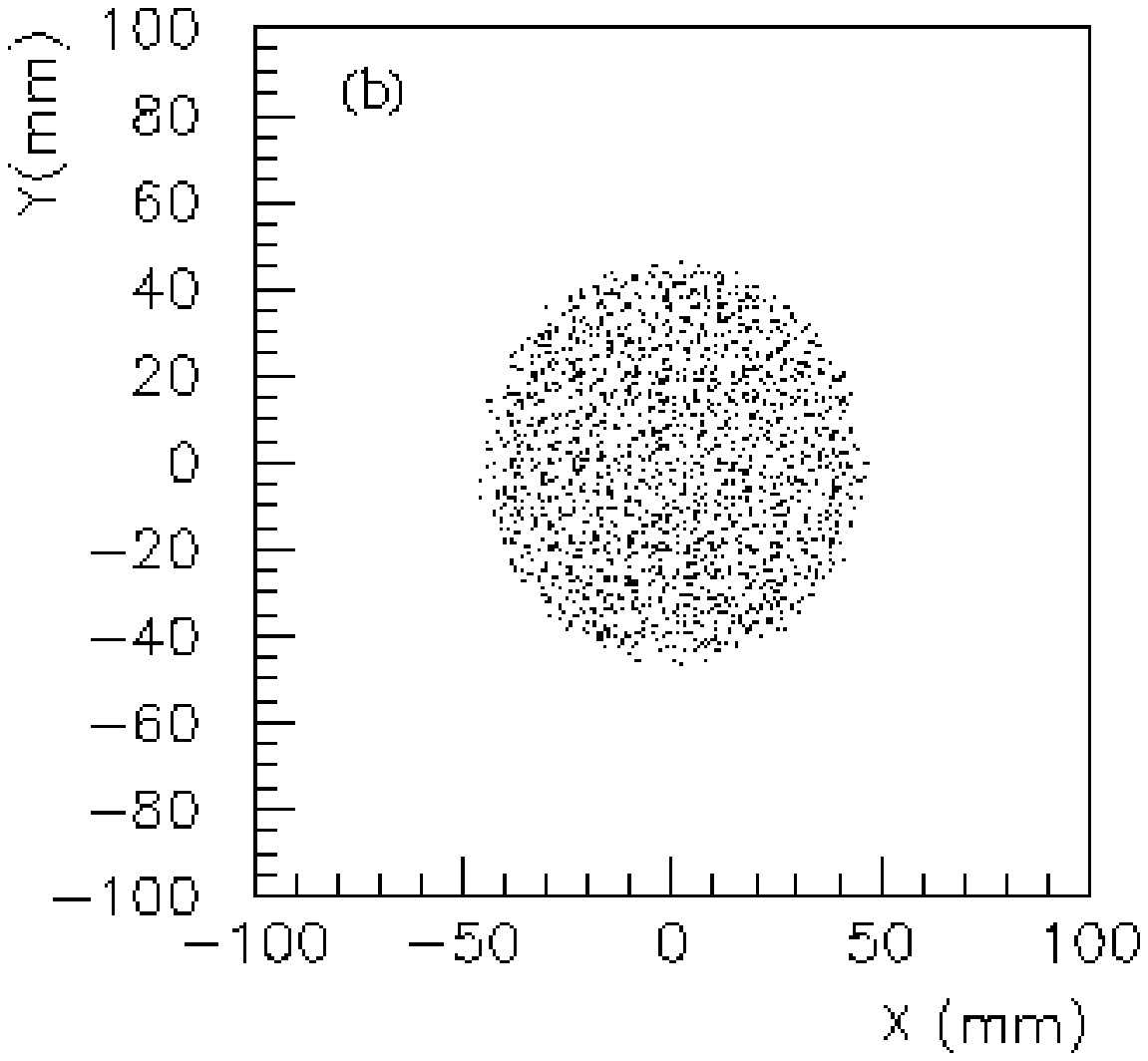}
\caption{\label{fig:epsart}(a) Neutron energy spectra resulting from a 
100 MeV proton beam on a 4 mm thick lithium target. (b) Scatter plot showing 
the
profile of the neutron beam at about 10 meters from the lithium target.}
\end{center}
\end{figure}
\subsection{MEDLEY set-up}

The first set-up downstream the beam was the MEDLEY detection array, described in
detail in reference~\cite{Da00}. 
Composed of eight Si-Si-CsI telescopes, this system is used to detect light charged 
particles up to $A=4$, with a low-energy threshold and over
an angular domain ranging from 20 to 160 degrees, in steps of 20 degrees. The 
telescopes were placed inside a vacuum reaction chamber of 100 cm diameter. 
The arrangement of the eight telescopes inside the chamber 
and a detailed view of one of them are given in Fig. 3. 
\begin{figure}
\begin{center}
\includegraphics[width=0.4\textwidth]{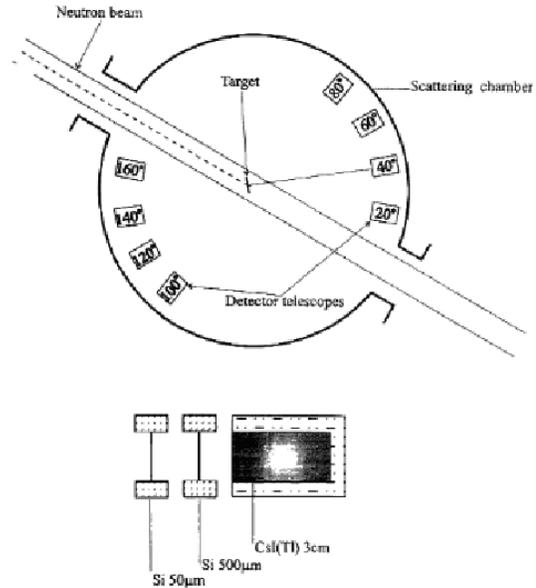}
\caption{\label{fig:epsart}MEDLEY detection array and the configuration of one
telescope.}
\end{center}
\end{figure}
For the MEDLEY set-up, the reaction target was placed at the center of the chamber
and was tilted
45 degrees with respect to the beam direction, in order to minimize the energy
loss of the produced particles inside the target. 
Typically, 50~$\mu$m thick targets were used for all experiments.
This allows small corrections for the energy loss of the
emitted particles inside the target, but it also results in a low particle production rate.
Due to the thin targets used and to the small solid angles of the telescopes,
the statistics accumulated using the MEDLEY set-up is relatively poor.
The angular resolution was defined by the target active area and by the opening 
angle subtended by each telescope. It has been estimated using Monte
Carlo simulations and typical values derived are of the
order of 5 degrees. 

\begin{figure}
\begin{center}
\includegraphics[width=0.4\textwidth]{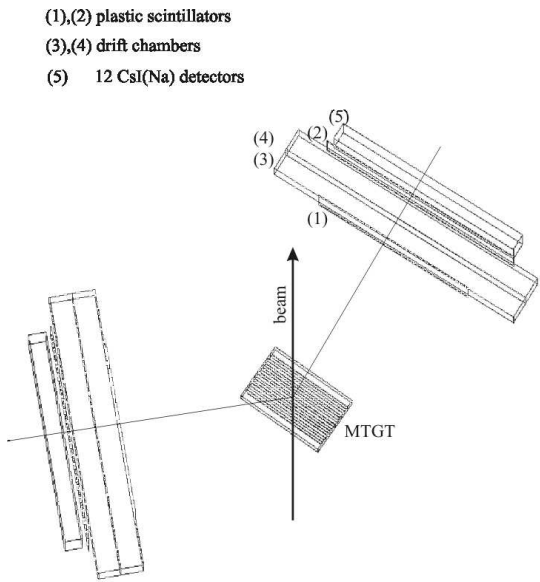}
\caption{\label{fig:epsart}Schematic view of SCANDAL set-up.}
\end{center}
\end{figure}

\subsection{SCANDAL set-up}
A detailed description of SCANDAL is given in reference~\cite{Kl02}. 
It consists of two identical systems located on each side of the neutron beam
and which covered a detection angular range of $10-140$ degrees (Fig. 4). 
Since particles travel in air before entering the set-up, 
only protons with energies larger than 30 MeV and a small number of 
deuterons could be detected. 
Each arm was composed of two 2 mm thick plastic scintillator used as triggers, two drift chambers
used for the particle tracking and an array of 12 CsI detectors enabling to 
measure particle residual energies. 
The emission angle of each particle was determined from its trajectory through the drift 
chambers. With this method, the angular resolution was estimated to be of the order of 1 degree,
which was a significant improvement compared to that obtained with the MEDLEY set-up. 
An example of an angular distribution obtained with particles detected in one 
of the CsI detectors is shown, together with simulation results, in Fig. 5.
The very good agreement observed demonstrates the validity of the tracking method used and
the quality of the drift chambers. Using the trajectories, 
the coordinates of the nuclear reactions on the target plane could be determined 
with a back-tracking procedure. Since the SCANDAL targets were larger than 
the neutron beam, it was crucial to determine the active target area with good precision.

\begin{figure}
\begin{center}
\includegraphics[width=0.3\textwidth]{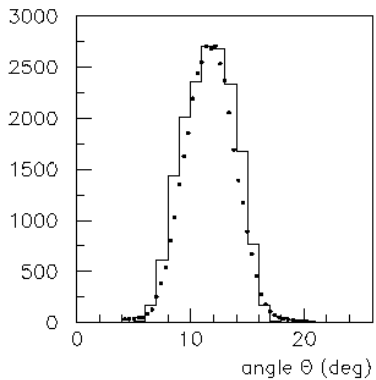}
\caption{\label{fig:epsart}Experimental distribution for emission 
angles of particles
detected in a CsI detector (dots) compared with the simulation 
results (histogram).}
\end{center}
\end{figure}

The SCANDAL set-up had the particularity to operate with a multitarget 
system (MTGT)~\cite{Co90}, which allows an increase of the counting rate
without impairing the energy resolution. Up to seven targets, inserted between 
multi-wire proportional counters (MWPC's), can be mounted simultaneously. 
The information given by MWPC's allows to determine the 
target from which the particle has been emitted, and to apply corrections to the particle energy
by taking into account the energy losses inside the subsequent targets.
In addition, by mounting simultaneously targets of different elements,
several nuclear reactions can be studied at the same time. During our experiments,
we operated with seven targets: five targets were made of the same material 
and dedicated to the reactions under study (iron, lead or uranium), another one
was a pure carbon target and the last one was a CH$_{2}$ target. 
By these means, events associated to the reactions under study and events
corresponding to the H(n,p) elastic scattering were recorded at the same time.
As will be explained in section III.C, those events enabled to apply an unambiguous 
normalization procedure for the extraction of the experimental cross sections, 
without requiring corrections for detection efficiencies, acquisition dead time
or beam intensity.

\section{Data reduction}
The data recorded using both detection systems were analyzed on a event-by 
event basis in order to extract the energy spectra of the emitted particles.
The procedures used for each set-up are described in the next two subsections.
The last subsection is dedicated to the cross section normalization method.

\subsection{Event sorting for SCANDAL set-up}
\begin{figure}
\begin{center}
\includegraphics[width=0.5\textwidth]{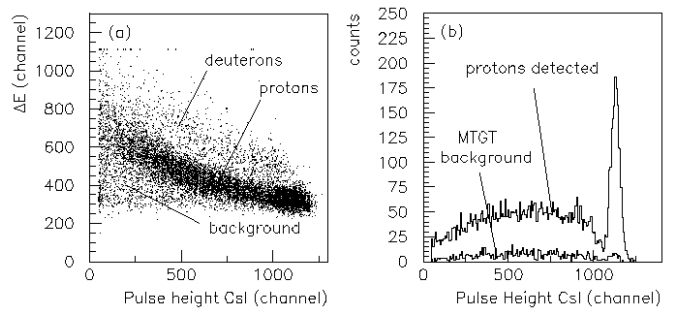}
\caption{\label{fig:epsart}(a) Two-dimensional scatter plot containing
events recorded in the 
$10-11$ degrees angular range using a CH$_{2}$ target. (b) Contamination in the recorded 
proton spectra due to reactions in the multitarget box.}
\end{center}
\end{figure}

\begin{figure}
\begin{center}
\includegraphics[width=0.5\textwidth]{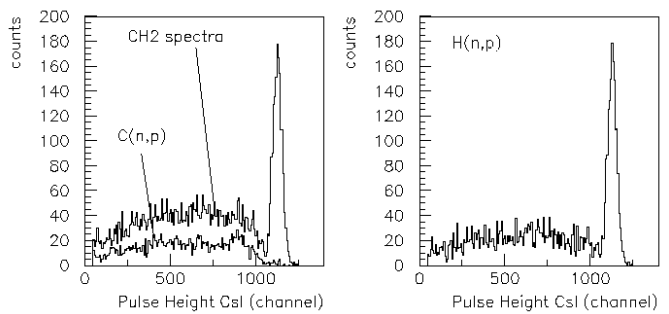}
\caption{\label{fig:epsart}Contribution of protons from the $^{12}$C(n,p) 
reaction in the CH$_{2}$
spectra (left part). On the right, the spectra of protons from 
the H(n,p) elastic 
scattering obtained after subtraction are shown.}
\end{center}
\end{figure}

The first step in the data analysis was to identify the target where the 
emission occured inside the multitarget system (MTGT). It was derived from
the signals given by the multi-wire proportional counters located in between the targets. 
Then the trajectories calculated with the drift chambers enable to determine the emission 
angle of each particle. In this way, both a target and an emission angle are associated to
each recorded event.

The particle identification was made by the well known $\Delta$E-E technique, using signals 
from the plastic scintillators and the CsI detectors. An example of such 
an indentification matrix is given in Fig. 6(a). It was obtained for the $10-11$ degrees
angular range, with a CH$_{2}$ target.
The small contribution of the deuterons which reached the CsI detectors, and a part of the 
background, were rejected by applying two-dimensional contours around the proton 
band. Another source of background which is present in the proton band, was due to 
protons from nuclear reactions that occured inside other multitarget box elements,
different from the targets of interest. Mainly, they were protons arising from $np$ scattering
reactions in the cathode foils. That additional pollution had to be rejected with another 
technique which consisted of recording "blank-target" events with the MTGT emptied of targets.
Subtraction of the corresponding spectra to those recorded during "physics" runs,
was performed after normalization to the same neutron fluency and to the same data acquisition
dead time. Examples of proton spectra  associated to blank-target runs and physics runs
are shown in Fig. 6(b).

With the CH$_{2}$ target, the energy calibration of the CsI detectors was done using protons
produced in H(n,p) reactions, for which the emission energies could be accurately calculated. 
In order to reject the contibution from $^{12}$C(n,p) reactions, a pure carbon target was 
mounted together with the CH$_{2}$ target inside the MTGT. Data on both 
targets were recorded simultaneously, so that, after normalization to the same number of 
carbon nuclei as in the CH$_{2}$ target, events associated to $^{12}$C(n,p) reactions
could be subtracted from the spectrum obtained with the CH$_{2}$ target.
Examples of spectra obtained with both targets are shown in Fig. 7, together with the
proton spectrum resulting from the subtraction. The latter presents a peak and a tail, reflecting the 
incident neutron spectrum presented in 
Fig. 2. Both features correspond to H(n,p) events induced, respectively, by 96 MeV
projectiles and by neutrons of lower energies contained in the beam tail. 
The proton energy spectra were obtained after calibration of the CsI detectors
and corrections for energy losses inside the set-up. These corrections were 
determined by Monte Carlo simulations for which attention has been paid to reproduce accurately 
the experimental conditions. The proton energy threshold equals 30 MeV. This large value is related
to the long flight (about 84 cm) through of air and detector materials of the system.

\subsection{Low-energy neutron rejection}

\begin{figure}
\begin{center}
\includegraphics[width=0.5\textwidth]{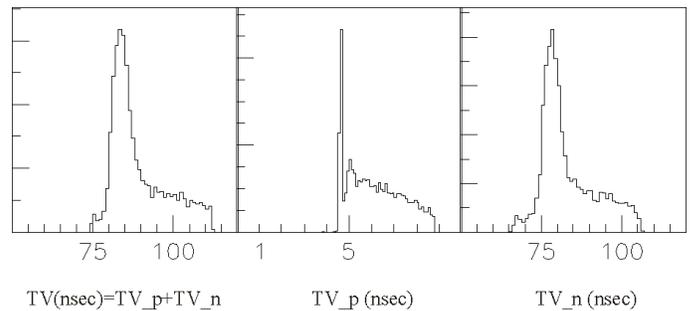}
\caption{\label{fig:epsart} Experimental determination of incident neutron time-of-flight.}
\end{center}
\end{figure}

In order to select only events induced by 96 MeV neutrons,
the contribution of low energy neutrons had to be rejected.
This has been done using a technique based on time-of-flight (tof). 
The tof measured were the sum of the neutron tof and the produced proton tof.
From the proton energy, the corresponding tof can be calculated and subtracted
from the total tof measured. The result is the tof of the neutron which induced the reaction.
In Fig. 8 are presented total time-of-flight, proton tof and neutron tof spectra.
The events associated to 96 MeV projectiles populate the peak centered at 78 nsec in the
neutron tof spectrum. This time corresponds to the  
experimental path of 1062.8 cm. A selection of that peak could then be easily applied.

In this way, spectra of protons from reactions induced by 96 MeV neutrons 
were constructed. In Fig. 9 two examples of such spectra obtained for Pb(n,Xp) 
and H(n,p) reactions recorded simultaneously with the MTGT system are presented.
As it can be seen, for H(n,p) elastic scattering reactions, after selection, only the 
peak at high energy remains in the spectrum, compared to that of Fig. 7, while the
contribution originating from low-energy neutrons has been completely removed.
This is a confidence check of the time-of-flight method used for 
the event selection. The statistics accumulated in both spectra presented in Fig. 9 
corresponds to about 2 hours of acquisition time. 

\begin{figure}
\begin{center}
\includegraphics[width=0.5\textwidth]{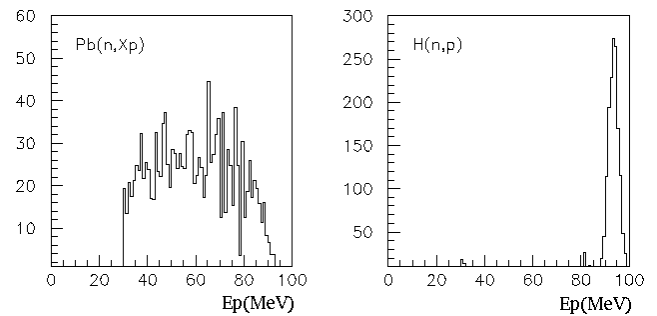}
\caption{\label{fig:epsart} Energy spectra of protons emitted in 
the angular range $10-11$ degrees 
from neutron-induced reactions on a lead target (left part) and 
from the elastic scattering
reaction (right part) at 96 MeV incident energy.}
\end{center}
\end{figure}

\subsection{Event sorting for MEDLEY set-up}
For the MEDLEY set-up, the particle identification has been done using the well known 
$\Delta$E-$\Delta$E and $\Delta$E-E techniques. Examples of two-dimensional plots obtained 
after energy calibration of each detector, for each particle type, are presented in Fig. 10.
The top figure represents particles stopped inside the second silicon detector, while the 
lower one shows particles which reached the CsI scintillator. 

\begin{figure}
\begin{center}
\includegraphics[width=0.5\textwidth]{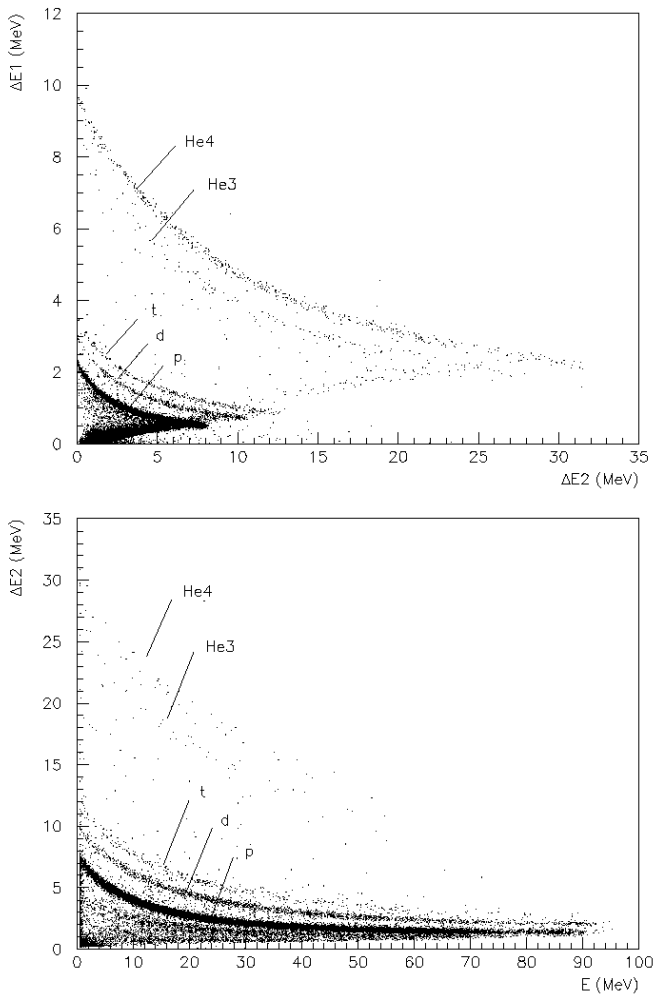}
\caption{\label{fig:epsart} Two-dimensional plots showing particles 
stopping in the second silicon
detector (top) and in the CsI detector (bottom) of the telescope placed 
at 40 degrees using a CH$_{2}$ target.}
\end{center}
\end{figure}

For calibration purposes, the points where each particle type start to punch through the 
silicon detectors were used. The corresponding energies were calculated 
with the detector thickness given by the manufacturer and the stopping power data from Ref. \cite{Zi85}. 
In addition, for the thin silicon detectors, the calibration was checked using 5.48 MeV alpha 
particles which stopped inside these detectors and which were emitted by a $^{241}$Am source.
The energy deposited in the CsI(Tl) detectors has been further calculated for each particle type 
using the energy losses
inside the silicon detectors. Supplementary calibration points in the case of protons were
provided by the H(n,p) reactions on a CH$_{2}$ target.
The method and the different parameterizations used are presented 
in detail in Ref. \cite{Da00}.

Finally, the total energy of each emitted particle is deduced by adding the different energies 
deposited inside the three individual detectors of each telescope. Fig. 11 shows energy 
spectra of p, d, t and alpha particles obtained from a lead target with the telescope placed
at 40 degrees. The arrows show the overlapping region between the 
second silicon and the CsI detector contributions.

\begin{figure}
\begin{center}
\includegraphics[width=0.5\textwidth]{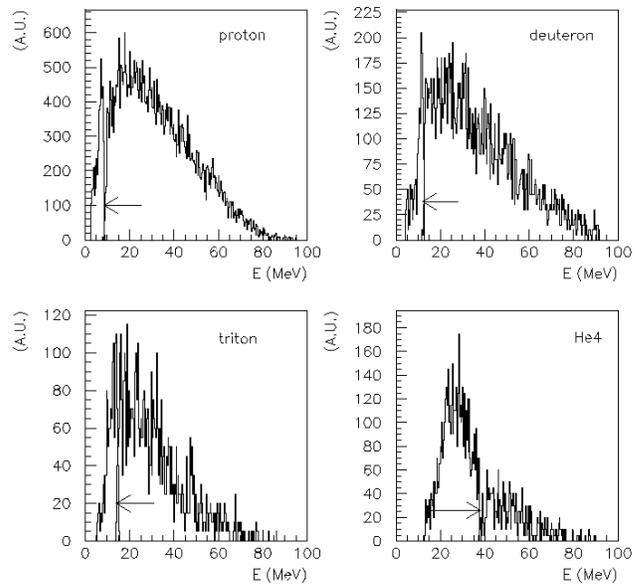}
\caption{\label{fig:epsart} Energy spectra for particles detected 
by the telescope placed at 40
degrees with all neutrons from the beam incident on a lead target.}
\end{center}
\end{figure}

The detection thresholds are given by the thickness of the first silicon detector. It is 
about $2-3$ MeV for the hydrogen isotopes and about 9 MeV for the helium isotopes, 
as it can be seen in Fig. 10.
The spectra had to be further corrected for the particle energy loss inside the emission target. 
Those corrections were calculated using Monte Carlo simulations,  with targets of about 50~$\mu$m
thickness. The maximum correction value estimated is less than 4 MeV, for low-energy alpha particles.
This shows that the corrections to be applied remain within reasonable limits. 

The rejection of events associated to low-energy neutrons was done with the same procedure as 
for SCANDAL (subsection B). 
The background is dominated by protons arising from neutron-induced reactions inside
the beam tube, at the entrance of the vacuum chamber.
That contribution is subtracted by using the spectra accumulated during blank-target runs,
applying a normalization to the same neutron fluency as for target-in runs, 
and taking into account corrections for the data acquisition dead time differences.

For the MEDLEY and SCANDAL set-ups, the CsI scintillator efficiency depends on the energy and 
type of the detected particle. Small corrections for the loss of light in the CsI detectors 
have then also to be applied. This effect is due to nuclear reactions which charged particles
can undergo while slowing down inside the CsI. Corrections for this effect have been estimated 
for all charged particles, using reaction cross sections available in the GEANT 
library from CERNLIB~\cite{Gea93}. Those estimations enable to determine the CsI detector efficiency 
as presented in Fig. 12 for protons (continuous line). 
The loss of light inside the CsI detector is rather important
for high energy protons and it is less pronounced for heavier particles. The 
detection efficiency at 100 MeV equals 91\% for protons, 93\% for deuterons, 95\% for 
tritons and 99\% for alpha particles and it increases as the energy decreases.
As shown in the figure, simulation results are in very good agreement with the
experimental values from reference~\cite{Kl02}.

\begin{figure}
\begin{center}
\includegraphics[width=0.3\textwidth]{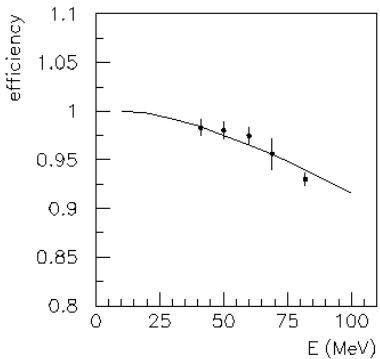}
\caption{\label{fig:epsart} Energy dependence of the CsI detection
efficiency for protons. Simulation result (continuous line) is compared 
with the experimental values from Ref. \cite{Kl02}.}
\end{center}
\end{figure}

\subsection{Cross section normalization}
Due to the difficulty encountered when monitoring a neutron
beam intensity, the absolute cross section normalization in neutron induced reactions is
a notorious problem. In particular for our experiments, the uncertainty
affecting the value given by the fission monitor equals 10 \%, which induces
large uncertainties for the values of the measured cross sections. 
Therefore, the cross sections are measured relatively to another one, considered as a reference.
The reference cross section the most often used is the H(n,p)
cross section, for which a recent measurement claims an absolute uncertainty of 2 \%~\cite{Ra01}.
We have used the values given in that reference
to calculate the absolute cross sections. Nevertheless, in order to be able to apply the
normalisation procedure, we have to measure in the same experimental conditions the number of 
protons emitted in H(n,p) reactions because that number intervenes in the 
normalisation factor. When measuring that number, we took the opportunity to remeasure the      
angular distribution of the H(n,p) cross section. 

For that purpose, we used the SCANDAL set-up. We determined the number 
of recoiling protons after subtraction of the $^{12}$C(n,p) reaction component and the background
contribution, following the procedure presented in subsection A. 
The angular range being limited in our measurements to $80-160$ degrees for neutrons in the center 
of mass system, we extracted values for the other angles by fitting our data with a fourth order
Legendre polynomial. Then, considering other channels 
negligible at 96 MeV, we normalized the value of the deduced total \textit{np} cross section
to that given in Ref. \cite{Li82}. Finally, we obtained the angular
distribution which is presented in Fig. 13 together with the experimental results of Ref. \cite{Ra01}.
We observe a very good agreement between both. However, 
the uncertainties of the cross sections from \cite{Ra01} are 
significantly smaller than in our experiment (2 \%).  
Indeed, for our data, the statistical errors are typically in the range $1.5-2.8$ \%, 
and the total uncertainty
is estimated of the order of 4.1 \%, including the 1 \% contribution 
from the total \textit{np} cross section \cite{Li82}. The
systematical errors affecting our results arise from the subtraction of reactions on carbon, 
from the integration over the \textit{np} peak and from the rejection of 
events induced by low-energy neutrons. 

\begin{figure}
\begin{center}
\includegraphics[width=0.3\textwidth]{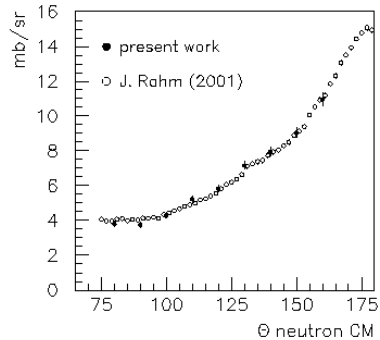}
\caption{\label{fig:epsart}Differential \textit{np} scattering cross section
at 96 MeV. The results obtained in the present work using SCANDAL set-up 
are compared to the data from reference \cite{Ra01}.}
\end{center}
\end{figure}

For the SCANDAL set-up, the MTGT system was used to measure at the same time protons 
emitted from the target under study (iron, lead or uranium) and in H(n,p) reaction 
from the CH$_{2}$ target. The normalisation procedure could then be applied without precise 
knowledge of the neutron flux.
For the MEDLEY set-up, all data have been normalized using the H(n,p)
scattering peak recorded by the telescope placed at 20 degrees during separate runs with 
a CH$_{2}$ target.

For the proton emission, data recorded using the SCANDAL and MEDLEY set-ups were individually 
normalized, allowing two independent determinations of the cross sections for
all targets studied.
With this procedure, the estimated systematical uncertainties of the experimental 
cross sections are not greater than 5.1 \%. To calculate this value, we took into account the 
contributions from the number of target nuclei (2 \%), the solid angles calculated by 
simulations (0.75 \%), the beam monitor stability during the data taking (2 \%), the
number of recoiling protons from the \textit{np} reaction (3.7 \%) and the reference \textit{np}
cross sections (2 \%) according to \cite{Ra01}.

\section{Experimental results}
The double-differential cross sections of light charged particles were measured for
three targets, Fe, Pb, U, with natural isotopic compositions, over an angular range of $20-160$ 
degrees. 
For the MEDLEY set-up, the low energy threshold was 4 MeV for hydrogen isotopes, 12 MeV for $^3$He and 
8 MeV for alpha particles. For the SCANDAL set-up, it was 35 MeV for protons. 
Due to the detector energy resolution and the available accumulated statistics, a 4 MeV bin size 
has been choosen for the energy spectra.

In the left part of Fig. 14, proton double differential cross sections 
measured independently with the MEDLEY set-up (full circles) and with the SCANDAL set-up
(open circles) are compared. The spectra correspond to the Fe target and a 20 degrees emission angle. 
Over the energy range covered by both detection devices, we observe a very good agreement. This
shows that the systematical uncertainties induced by the cross section normalization
are small. We obtained similar results for the other targets (Pb and U) and over the full
angular range. In addition, it shows that the limited angular resolution of MEDLEY
does not distort the distributions which are comparable to that obtained with SCANDAL, for which 
the angular resolution is much better.
The right part of Fig. 14 gives a comparison of the Fe(n,Xp) cross section measured at 20
degrees with MEDLEY, with the data from Ref. \cite{Ron93} which were obtained
using the magnetic spectrometer LISA. A very good agreement is found also between these two measurements.
This shows the quality of our measurements and of the data analysis procedures employed.
We observed a similar agreement for the Pb(n,Xp) cross sections.

\begin{figure}
\begin{center}
\includegraphics[width=0.5\textwidth]{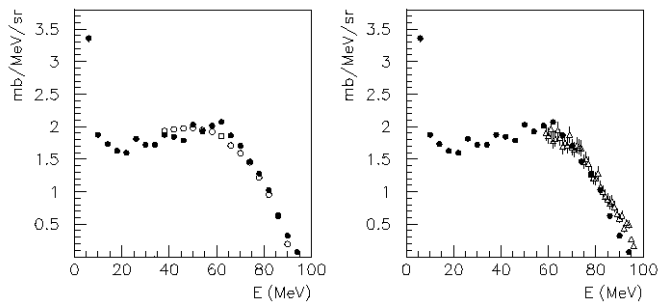}
\caption{\label{fig:epsart} Left panel: Fe(n,Xp) double-differential cross sections measured with 
MEDLEY set-up at $\theta$=20 degrees (full circles), compared to the SCANDAL results (open circles).
Right panel: same data compared to those from Ref. \cite{Ron93} (open triangles).}
\end{center}
\end{figure}

In Fig. 15 are presented experimental double-differential cross sections for p, d, t 
(top, middle, bottom lines respectively) produced in 96 MeV neutron induced reactions on 
Fe, Pb and U targets
(left, middle and right rows respectively) and measured with the MEDLEY set-up. In Fig. 16, 
for the same reactions, we report the complementary production cross sections of $^3$He and alpha 
particles (top and bottom figures respectively). The errors shown are purely statistical.

\begin{figure*}
\begin{center}
\includegraphics[width=1.0\textwidth,height=1.0\textheight]{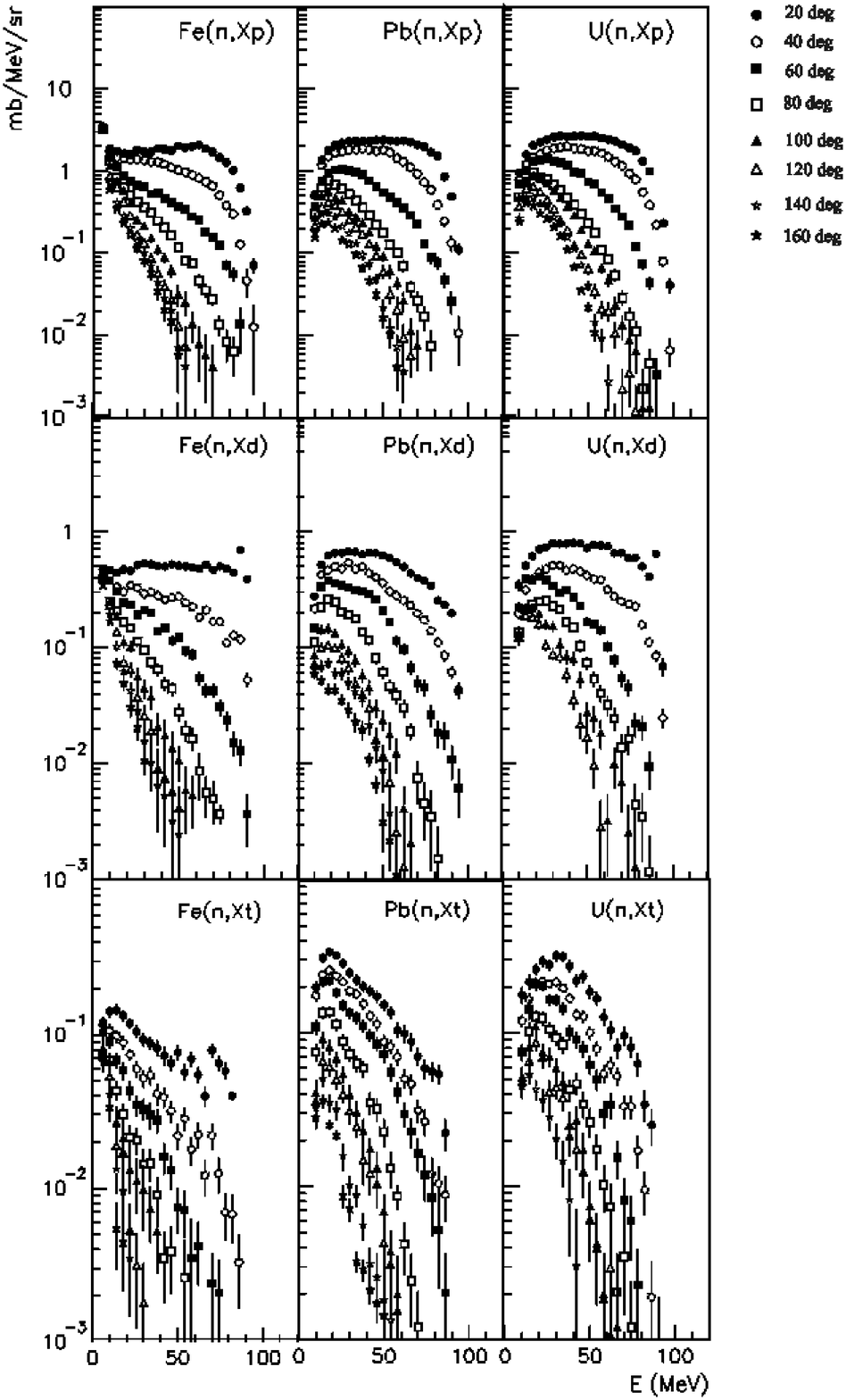}
\caption{\label{fig:epsart} Double-differential cross sections for p, d, t particles
(top, middle and bottom lines respectively) emitted in 96 MeV neutron-induced reactions
 on Fe, Pb and U targets (left, middle, right rows respectively).}
\end{center}
\end{figure*}

\begin{figure*}
\begin{center}
\includegraphics[width=0.9\textwidth,height=0.6\textheight]{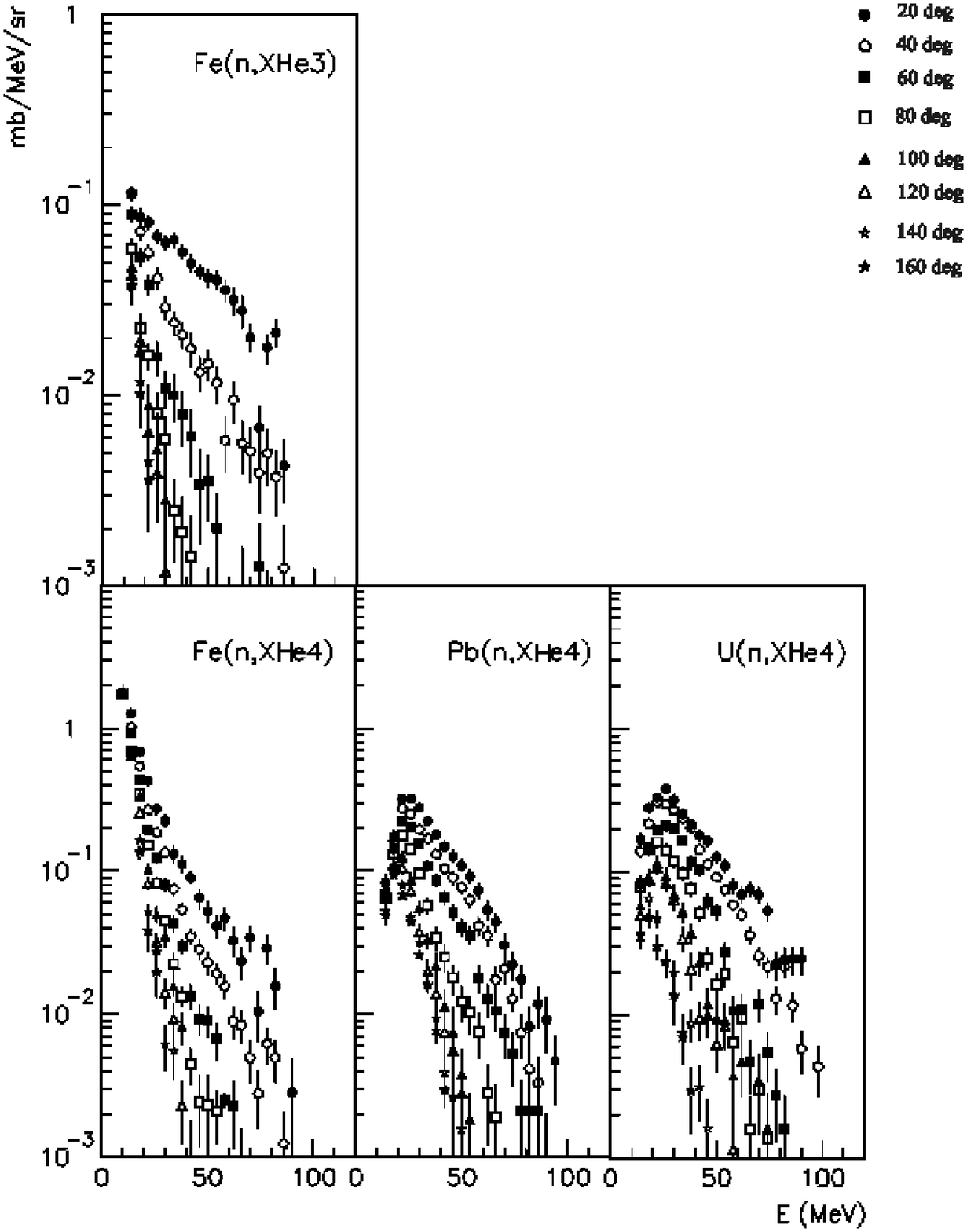}
\caption{\label{fig:epsart} Same as Fig. 15 for helium isotopes.}
\end{center}
\end{figure*}

The general trend observed is a decreasing emission probability with increasing angle, over the full 
energy range. The angular distributions are slightly forward peaked at low energies, and at backward
angles, the emission probabilities are very low for energetic particles.
In the case of the iron target, a quasi-isotropic component is observed at very low 
energy ($0-10$ MeV). This contribution is not present for heavier targets, for which Coulomb
effects are much larger. For the rest of the energy range, the distributions obtained with the 
three targets are very similar in shape. 
For $^3$He particles, distributions have been measured only for the iron target. 
Despite the long data acquisition time, no corresponding events were recorded for the other targets.
This is related to the very low $^3$He emission probability for heavy targets which has been
already observed in Ref. \cite{Gue04}, where the $^3$He production rate in 63 MeV proton 
induced reactions on $^{208}$Pb is about 10 times smaller than for tritons.

\begin{figure}
\begin{center}
\includegraphics[width=0.5\textwidth]{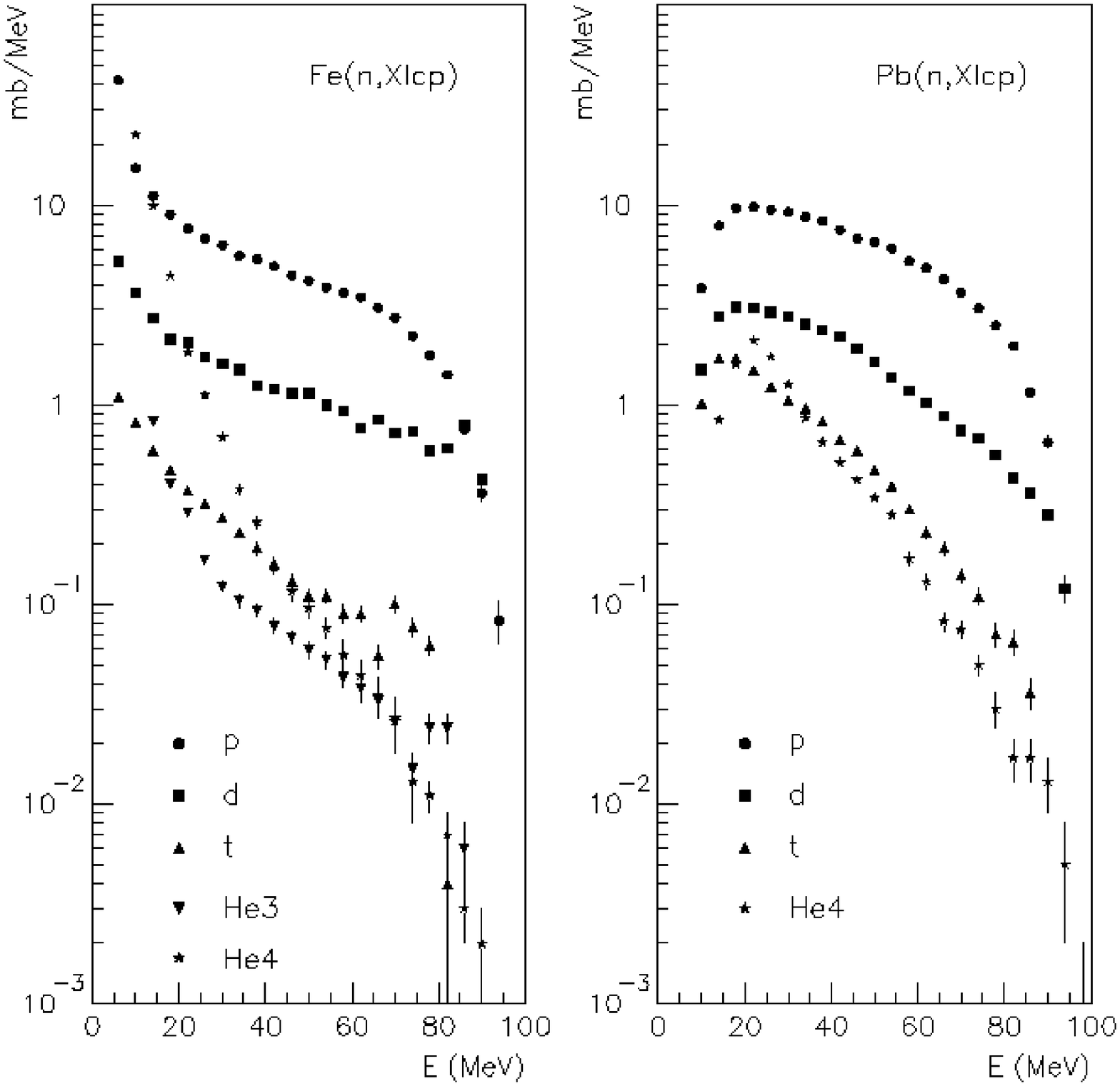}
\caption{\label{fig:epsart} Energy distributions for light charged
particle produced in 96 MeV neutron induced reaction on iron and lead 
targets (left and right panels respectively).}
\end{center}
\end{figure}

\begin{figure}
\begin{center}
\includegraphics[width=0.5\textwidth]{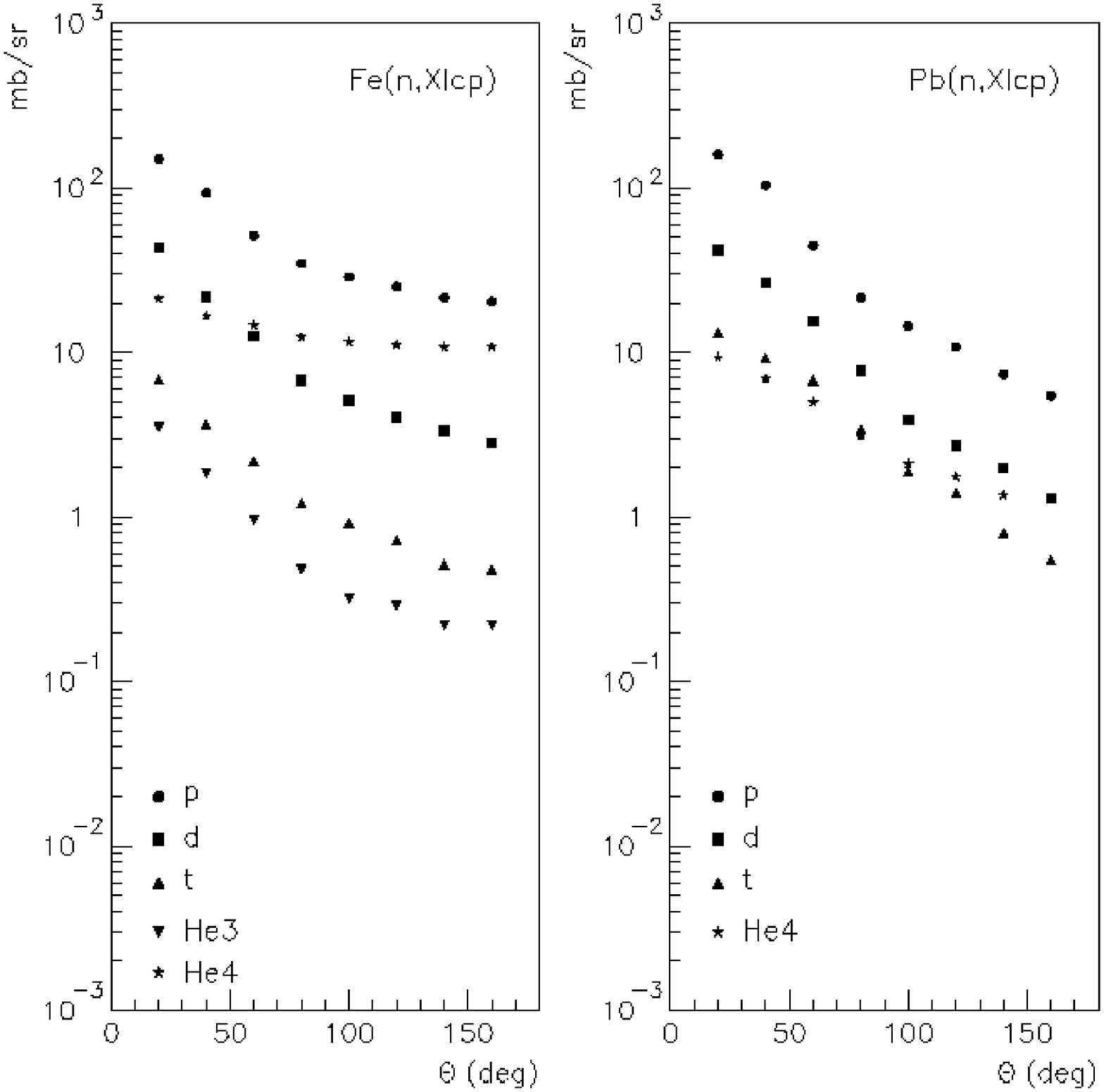}
\caption{\label{fig:epsart}Angular distributions for light charged
particle produced in 96 MeV neutron induced reaction on iron and lead 
targets (left and right panels respectively).}
\end{center}
\end{figure}

For a more detailed analysis of the particle emission mechanisms, a separate inspection of angular and
energy differential distributions is needed. The angular distributions were obtained from 
double-differential cross sections by integration over the full energy range.
For the energy distributions, we used the Kalbach systematics~\cite{Kal88} in order 
to extrapolate the experimentally available angular range over the entire space. Finally, the
total production cross sections were derived for each particle type by integrating the
corresponding energy-differential cross sections over the experimental energy range.

The energy-differential cross sections are presented in Fig. 17 for the iron and lead
targets. The results obtained with the uranium target are very similar to those extracted for
the lead one. By analyzing the spectra, we distinguish 
two regions. For energies greater than about 20 MeV, proton and 
deuteron spectra are very similar in shape, the emission probability decreasing slowly with
energy for both targets. 
In the case of the iron target, the triton and $^3$He spectra also show a similar 
behavior. 
For alpha particles, the spectra decrease very rapidly with energy. 
For a given particle, the shapes of the iron and lead distributions are very
similar.
In this energy region, the emission probability distributions have steeper slopes 
for heavy particles than for light ones.
Another important aspect to be noticed is the decreasing emission probability with the nucleon 
number of the emitted particle. However, an exception is observed for alpha particles
for which the production cross sections in the low-energy part of the continuum region 
(20 MeV$<$E$<$35 MeV) are larger than for tritons, suggesting a more complex mechanism for their 
emission.
For low emission energies (E$<$20 MeV), a dominant contribution is observed for all particles 
in the case of the iron target. The shape of the distributions in Fig. 17, correlated to the 
slow variation of the amplitude with the emission angle observed  
in Figs. 15 and 16, suggests that these low energy particles are emitted mainly 
during the evaporation process of an excited nucleus. This component is not present in
the spectra obtained with the lead and uranium targets because, for heavy targets, the emission of
low energy particles is strongly inhibited by Coulomb effects. This explains the
low cross sections found in this energy range for both lead and uranium targets.

Fig. 18 shows the angular distributions obtained by integrating double-differential 
distributions. Due to the detection thresholds, the integration domains range over 
$4-96$ MeV for hydrogen isotopes, $12-96$ MeV for $^3$He and $8-96$ MeV for alpha particles.
For all particles, the distributions are strongly forward peaked, suggesting that non-equilibrium
processes are dominant for the reactions under study. An
exception can be noticed for alpha particles in Fe(n,X) reactions, where the
distribution is almost flat for angles larger than 50 degrees. This suggests that the emission of
alpha particles in the backward hemisphere could result mainly from evaporation processes. 
For a given particle, the angular distributions are more forward peaked for the heavier nuclei, 
suggesting that the non-equilibrium component increases with the nucleon number of the
target. 

\begin{table*}
\begin{center}
\begin{tabular}{|c|c|c|}
\hline
\textbf{Reaction}&
\textbf{Total cross section}&\textbf{Non-equilibrium cross section} \\
\textbf{}&\textbf{(mb)}&\textbf{(mb)}
\\ \hline
Fe(n,Xp)&584$\pm$29.2&326
\\
Pb(n,Xp)&485$\pm$24.3&485
\\
U(n,Xp)&589$\pm$29.5&589
\\ \hline
Fe(n,Xd)&131$\pm$6.5&96
\\
Pb(n,Xd)&137$\pm$6.9&137
\\
U(n,Xd)&170$\pm$8.5&170
\\ \hline
Fe(n,Xt)&21$\pm$1.1&15
\\
Pb(n,Xt)&53$\pm$2.7&53
\\
U(n,Xt)&54$\pm$2.8&54
\\ \hline
Fe(n,X$^3$He)&10$\pm$0.5&7
\\ \hline
Fe(n,X$^4$He)&167$\pm$8.3&31
\\
Pb(n,X$^4$He)&45$\pm$2.2&45
\\
U(n,X$^4$He)&52$\pm$2.6&52 \\
\hline
\end{tabular}
\end{center}
\caption{Total light charged particle integral cross sections and 
estimated contributions from the non-equilibrium emission in neutron 
induced reactions at 96 MeV.}
\end{table*}

This can also be observed from Table I, where integrated total cross sections (second column) and 
integrated non-equilibrium cross sections (third column) are presented as a function of the target mass,
for all particles. Depending on the system considered, the non-equilibrium cross sections were
extracted with different methods. 
For the Fe(n,Xlcp) reactions (lcp refers to light charged particles), the low-energy contribution in the 
energy-differential cross sections (Fig. 17) was fitted with 
an exponential function and its integral was then subtracted from the total 
cross section for each particle. For lead and uranium targets, we made the assumption
that all particles were emitted during non-equilibrium processes, i.e., in a first approximation,
the rather small contribution of evaporated particles expected at low energy 
is neglected. 

The values from Table I show that for all 
targets studied, more than 30\% of the total light charged particle production are 
particles heavier than protons. This is an important aspect to be pointed out, because, with such 
a production rate, the contribution of these particles should not be neglected.

\section{Theoretical calculations}
At this moment, the exciton model \cite{Gr66} is the most commonly used to calculate the 
pre-equilibrium emission in nucleon-induced reactions at intermediate energies. This model 
assumes that the excitation process takes place by successive nucleon-nucleon interactions
inside the nucleus. Each interaction produces another exciton, leading the system to the final
state of statistical equilibrium through more complex states. Occasionally a
particle can receive enough energy to leave the system and subsequently be emitted. The
resulting pre-equilibrium spectrum is the sum of the contribution from each state. Particles
emitted in the early stages have more energy than those emitted in the later ones.
In the framework of this model, only energy distributions of emitted particles can be calculated.

The GNASH code \cite{Ch92} is one example which uses the exciton model to calculate the 
pre-equilibrium component. It is able to calculate spectra for nucleons and complex particles. 
In this code, the equilibrium contribution is calculated using the Hauser-Feshbach formalism
\cite{Ha52}. Cross sections which were evaluated with GNASH are at present implemented
in MCNPX, a code widely used for specific applications such as medical or engineering studies. 
In Figs. 19 and 20, we compare, respectively, the $^{56}$Fe(n,Xlcp) and $^{208}$Pb(n,Xlcp) 
energy-differential cross sections of the present work (points), to 
the GNASH predictions (histograms) obtained with MCNPX.
The maximum value in the alpha particle spectrum for the iron target has been set to 1 mb/MeV for 
a better visualisation.
While the proton emission is relatively well described, we observe that the production of
complex particles is strongly underestimated.

\begin{figure}
\begin{center}
\includegraphics[width=0.5\textwidth]{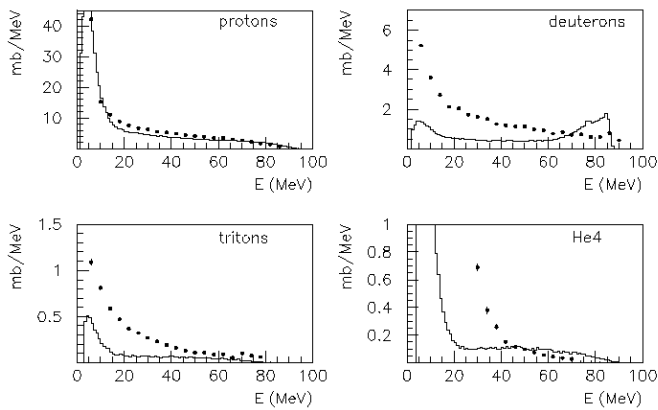}
\caption{\label{fig:epsart}Energy-differential cross sections calculated using the GNASH code 
for the $^{56}$Fe(n,Xlcp) reaction at 96 MeV (histograms) compared with the present experimental results 
(points).}
\end{center}
\end{figure}

\begin{figure}
\begin{center}
\includegraphics[width=0.5\textwidth]{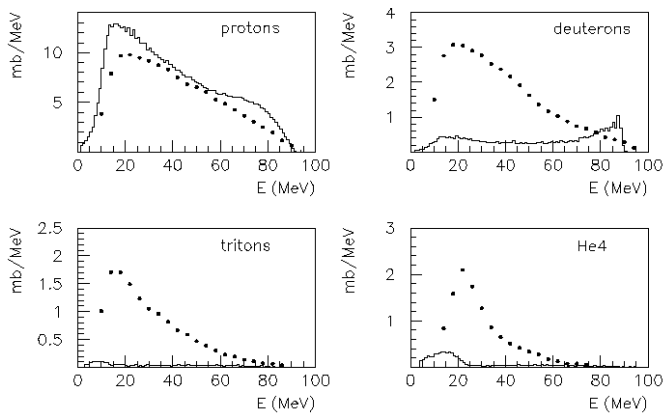}
\caption{\label{fig:epsart}Same as in Fig. 19 for the $^{208}$Pb(n,Xlcp) reaction.}
\end{center}
\end{figure}

This comparison suggests that significant improvements are needed in the original exciton 
model in order to increase its prediction level concerning the cluster emission. 
To modify it according to this request, a first approach was proposed in 1973 by 
Ribansk\'y and Oblo\v zinsk\'y~\cite{Ri73}. It introduces the probability of a cluster formation
during the nucleon-nucleon interactions inside the target. 
In 1977, Kalbach formulated a second approach~\cite{Kal77} which includes contributions from direct 
pick-up and knock-out mechanisms. Both approaches have been tested in the past 
against data and they lead to a satisfactory agreement with the experimental results 
~\cite{Kal77,Wu78}, despite their completely different basic assumptions. 
Nevertheless, conclusions about their global predictive power were limited, 
mainly because a restricted number of experimental results were available at that moment.
In order to get a wider view on their predictive capabilities, we performed calculations with both
approaches for the $^{56}$Fe(n,Xlcp), $^{208}$Pb(n,Xlcp), $^{238}$U(n,Xlcp) reactions at 96 MeV, but also 
for other projectiles, at different incident energies and for other targets, for which experimental data
are available in the literature. In the forthcoming subsections, we will give a basic description of both
approaches and discuss the comparaisons of the calculations with a set of data that  
cover a wide domain of reaction entrance-channel parameters.

\subsection{Cluster formation probability in nucleon-nucleus reactions}
\begin{figure*}
\begin{center}
\includegraphics[width=0.8\textwidth]{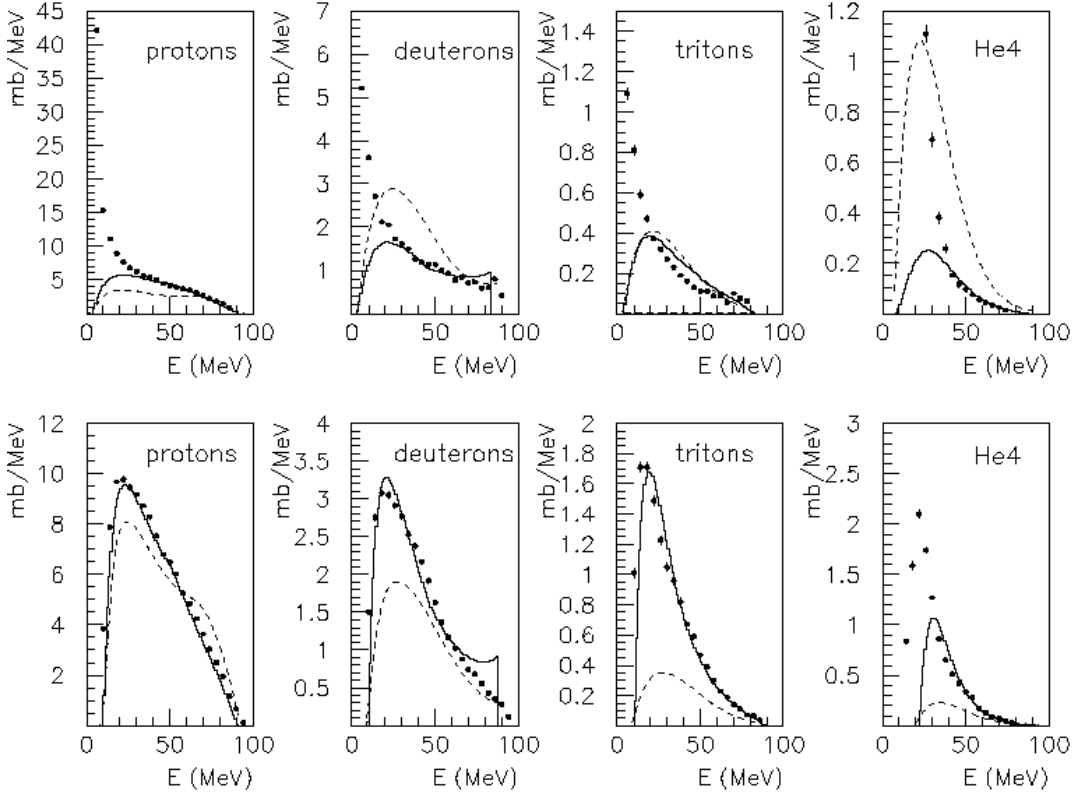}
\caption{\label{fig:epsart}Energy-differential cross sections calculated using PREEQ (histograms)
and PRECO-2000 (dashed lines)
for $^{56}$Fe(n,Xlcp) (top) and $^{208}$Pb(n,Xlcp) (bottom) reactions at 96 MeV,
compared with the experimental results of the present work
(points). Maximum value in the alpha-particle spectrum in the case of the iron target 
has been set to 1.2 mb/MeV for a better comparison.}
\end{center}
\end{figure*}

Due to the difficulties encountered by the original exciton model proposed by Griffin
to reproduce the experimental distributions of complex particles, 
it has first been modified by 
Ribansk\'y and Oblo\v zinsk\'y~\cite{Ri73}. The modification consists of the introduction 
in the particle production rate expression of a multiplicative term containing the cluster
formation probability $\gamma$$_{\beta}$ where $\beta$ is the type of the emitted particle.
The physical meaning of this parameter has been given in Ref. \cite{Mac79} in the framework
of the coalescence model. This approach assumes that complex particles are formed during the
pre-equilibrium stage from excited nucleons which share the same volume in the momentum 
space.
In this way, for example, an excited proton and an excited neutron can coalesce into a 
deuteron if the transverse momentum between both is small. 
The drawback of this approach is its limited predictive power since the parameter 
$\gamma$$_{\beta}$ has to be adjusted in order to reproduce as well as possible the amplitude
of the experimental energy-differential distribution under study. Nevertheless, 
it is always interesting to compare the tuned results of a model with experimental data.

\begin{figure*}
\begin{center}
\includegraphics[width=0.8\textwidth]{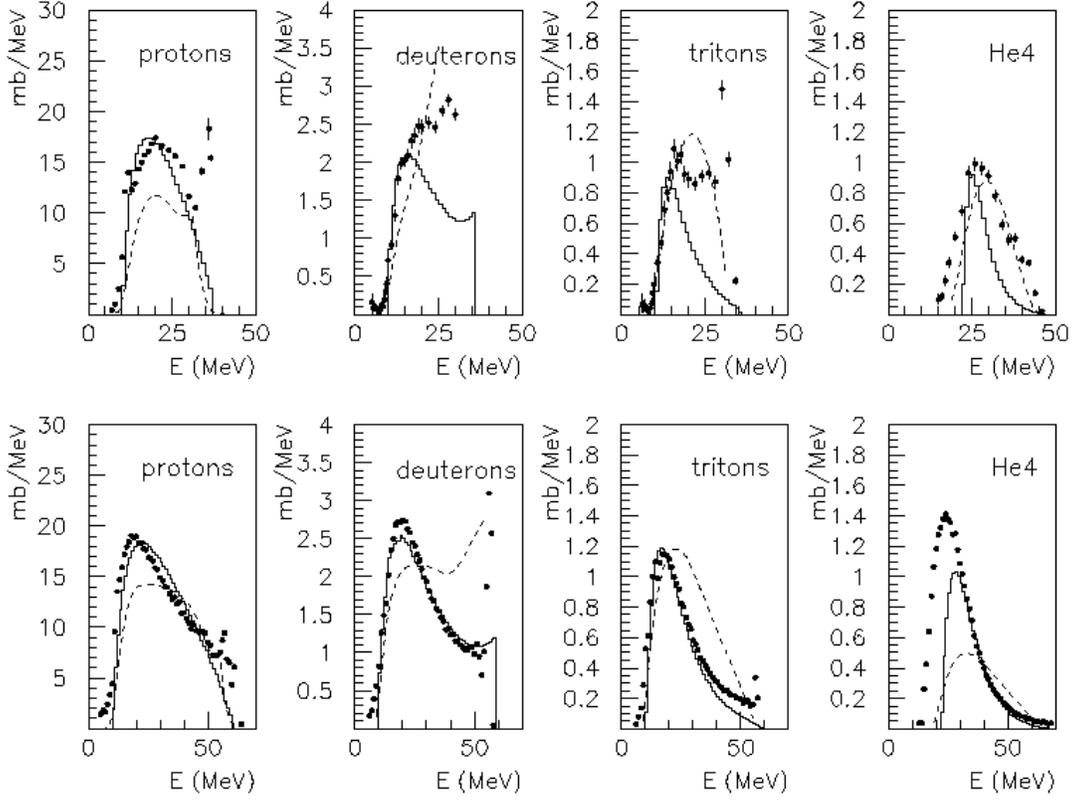}
\caption{\label{fig:epsart}Energy-differential cross sections calculated using PREEQ (histograms)
and PRECO-2000 (dashed lines) for $^{209}$Bi(p,Xlcp) reactions at 39 MeV (top) and $^{208}$Pb(p,Xlcp) 
reactions at 63 MeV (bottom), compared with the experimental results from 
Ref. \cite{Ber73} and \cite{Gue04} (points).}
\end{center}
\end{figure*}
The formation probability $\gamma$$_{\beta}$ of a complex particle $\beta$ is given as 
a function of the radius of the coalescence sphere $P$$_{0}$ in the momentum space by the formula:
\begin{equation}
\gamma_\beta=|\frac{4}{3}\pi(P_0/mc)^3|^{p_\beta-1}
\end{equation}
where $p$$_\beta$ is the number of nucleons of the emitted particle. 
Of course, $\gamma$$_{\beta}$=1 in the case of the emission of a nucleon. 
According to equation (1),  $\gamma$$_{\beta}$ and thus, $P$$_{0}$ is the free 
parameter of the model. 

The following expression for the cluster formation probability has been proposed
in the Ref. \cite{Gud83}:
\begin{equation}
\gamma_\beta=(p_\beta)^3(p_\beta/A)^{p_\beta-1}
\end{equation}
where $A$ is the mass of the target nucleus. This approach is implemented in the latest version 
of the code GEANT \cite{Gea03}, which is intensively used for simulations among the physics
community. However, calculations from Ref. \cite{Gud83} strongly overestimates deuteron, triton and 
$^3$He
distributions, while the production rates for alpha particles are underestimated. This shows that the
calculation of the cluster formation probability according to equation (2) is not very 
appropriate.
For this reason, calculations in this work have been done with the PREEQ program \cite{Bet75},
keeping the cluster formation probability as a free parameter.
A complete explanation about the different parameters of the model and the method we applied
to calculate them can be found in Ref. \cite{Wu78}. In the forthcoming discussion,
we will focus onto the cluster formation probability $\gamma$$_{\beta}$ because of its
particular importance for the model predictions. 

\begin{figure*}
\begin{center}
\includegraphics[width=0.8\textwidth]{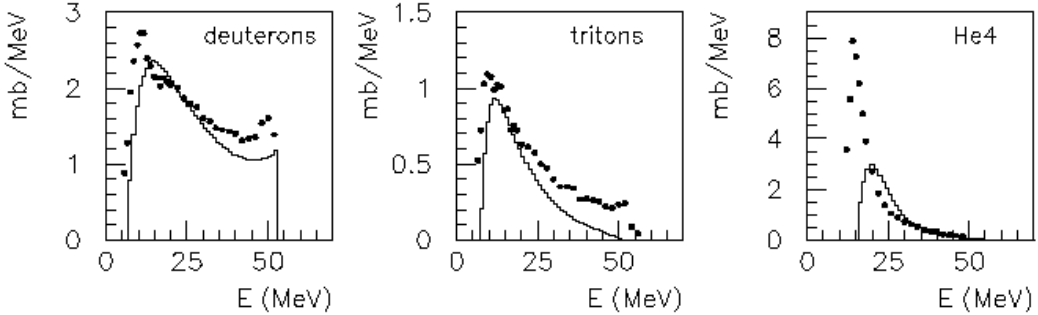}
\caption{\label{fig:epsart}Energy-differential cross sections calculated using the PREEQ code
(histograms) for $^{120}$Sn(p,Xlcp) reaction at 62 MeV compared with the experimental results from 
the Ref. \cite{Ber73} (points).}
\end{center}
\end{figure*}

In the first step of our investigation, we performed calculations with PREEQ 
for the 96 MeV neutron induced reactions presented in this work. 
We determined two sets of values for the  $\gamma$$_{\beta}$ parameter by normalizing 
individually the calculated energy distributions to the Fe(n,X) and 
Pb(n,X) experimental data. For those reactions, PREEQ results (histograms) and data (points)
are presented in Fig. 21 for $^{56}$Fe and $^{208}$Pb targets. We have to remind 
that the model calculates only the pre-equilibrium component of the emission spectra, so 
that in our comparison, we should not consider neither the low-energy region 
populated with particles evaporated by excited nuclei, nor the high-energy region
where direct reactions are supposed to be dominant. Considering those restrictions, we observe that
the shapes of the calculated distributions are in good agreement with the experimental results.
As expected, the model fails in reproducing the very low-energy component of the iron spectra.
For all particles in the $^{208}$Pb(n,X) reactions, except alpha particles, 
the calculated pre-equilibrium contribution accounts for almost the entire energy
range, showing that almost all particles are emitted during the pre-equilibrium stage.
For alpha particles, the pre-equilibrium processes are still underestimated by PREEQ in the 
low-energy region of the continuum. 
By comparison with the GNASH predictions presented in Figs. 19 and 20, we clearly see that
this approach improves dramatically the original exciton model, for all particles.

For protons, for which the $\gamma$$_{\beta}$ parameter equals 1 and has not to be adjusted,
the amplitudes of the distributions are very well described 
by the model in the energy range where it is applicable. It must be pointed out that only the
primary pre-equilibrium contribution is calculated by this code. The good agreement found
for protons suggests that the second-chance preequilibrium component is very small, in
agreement with the calculations from Ref. \cite{Bla76}.
For complex particles, no conclusion about the predictive capabilities of PREEQ can be drawn at the
moment, the amplitude of the distributions being obtained by adjusting the $\gamma$$_{\beta}$ parameter 
in order to get a good agreement with the experimental data. 
Therefore, the next step in our analysis was to check the stability of this parameter 
while changing the entrance channel, i.e., the incident energy and the projectile, for a target nucleus
in the mass region $A=208$.
For that aim, using the values of the cluster formation 
probabilities previously obtained for the 96 MeV $^{208}$Pb(n,X) reactions, we calculated the energy 
differential cross sections for 39 MeV $^{209}$Bi(p,X) and 63 MeV $^{208}$Pb(p,X) reactions. 
In Fig. 22, the resulting 
PREEQ calculations (histograms) are compared with the experimental data (points) measured at 39 MeV with a
$^{209}$Bi target \cite{Ber73} (top panels) and to the data measured at 63 MeV with the $^{208}$Pb target 
\cite{Gue04} (bottom panels).
Over the energy domain where the model is applicable, we observe again a good agreement between the 
calculations and the experimental results. In addition and especially at 39 MeV, we see that the
non-calculated direct process contribution is dominant. 

From this study, we conclude that the free parameter $\gamma$$_{\beta}$ depends neither on the 
projectile type (neutron or proton) nor on the incident energy and that, once the cluster formation 
probability has been adjusted on reaction system, the model has a good predictive power 
for reactions with the same target. To go further, we have now to investigate 
its possible dependence with the target mass. Since we have just determined the formation probability
$\gamma$$_{\beta}$ for two target nuclei with masses $A=56$ and $A=208$, we choose an intermediate 
target with a mass $A=120$ for which experimental data were measured, i.e., the $^{120}$Sn(p,Xlcp) reaction 
at 62 MeV incident energy \cite{Ber73}. With the same method described previously, we calculated the 
new set of $\gamma$$_{\beta}$ values associated to the $^{120}$Sn target. In Fig. 23, we compare the 
corresponding PREEQ calculations (histograms) to the data (points). As for the other targets, we observe 
the same global good reproduction of the data in the pre-equilibrium energy region.

\begin{table*}
\begin{center}
\begin{tabular}{|c|c|c|c|c|}
\hline
\textbf{Target}&\textbf{Emitted particle}&\textbf{Formation probability $\gamma$$_{\beta}$}&\textbf{$P$$_{0}$} \\
\textbf{}&\textbf{}&\textbf{}&\textbf{(MeV/c)}
\\ \hline
$^{56}$Fe
&  d & 0.0278 & 175 \\
&  t & 0.0065 & 250 \\
&  $^3$He & 0.0060 & 246 \\
&  $^4$He & 0.0052 & 322 \\
\hline
$^{120}$Sn
&  d & 0.0230 & 164 \\
&  t & 0.0050 & 238 \\
&  $^4$He & 0.0035 & 304 \\ 
\hline
$^{208}$Pb
&  d & 0.0186 & 153 \\
&  t & 0.0035 & 225 \\
&  $^4$He & 0.0018 & 286 \\ 
\hline
\end{tabular}
\end{center}
\caption{Cluster formation probability in nucleon-induced reactions on three targets and
corresponding radii of the coalescence sphere in the momentum space.}
\end{table*}

In Table II, we gather the values of the cluster formation probabilities, as well as the 
related $P$$_{0}$ parameters, obtained for the three target nuclei $A=56$, $A=120$ 
and $A=208$ and for each complex particle type. The formation probability
for each particle type is also  represented as a function of the target mass in Fig. 24.

\begin{figure}
\begin{center}
\includegraphics[width=0.3\textwidth]{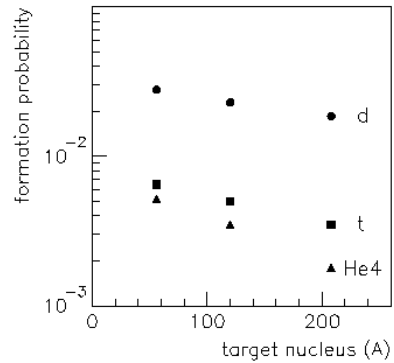}
\caption{\label{fig:epsart}Formation probability for each complex particle versus target mass.}
\end{center}
\end{figure}
We observe that for a given particle, the formation probability and then the
coalescence sphere radii, is smaller for heavier nuclei. Under the assumption of phase space
relations, a smaller $P$$_{0}$ value means a larger volume inside the nucleus from which the particle
is emitted. This volume is then larger for heavier nuclei. In addition, for a given target nucleus,
the figure shows that the formation probability decreases as the number of nucleons of the 
emitted particle increases. This could be explained by the fact that it
is less probable, for example, for three nucleons to coalesce in order to form a triton, than
for two nucleons to form a deuteron. The formation probability of deuterons is much larger
than for other complex particles, suggesting that the most probable
mechanism is the pick-up of one nucleon by another. 

The presently obtained values are in rather strong disagreement with those
from Ref. \cite{Gud83}. Thus, for the $^{208}$Pb(p,Xlcp) reaction, the $\gamma$$_{\beta}$ probability 
calculated 
according to equation (2) is 0.077 for deuterons, 0.0056 for tritons and 0.00046 for alpha particles.
As it can be observed, the values for hydrogen isotopes are larger than those obtained in this work, leading to
the overestimation found in the Ref. \cite{Gud83} for the production of these particles. On the other hand,
the values for alpha particles are smaller than ours and thus the distributions calculated in 
Ref. \cite{Gud83} are systematically below the experimental results.

Another interesting aspect to point out is that the presented $P$$_{0}$ values obtained for 
nucleon-nucleus reactions are in the same range as those extracted for reactions induced by 
complex projectiles (deuterons, $^3$He and alpha particles) at intermediate energies \cite{Mac80} and
for reactions induced by heavy ions at high energies \cite{Awe81,Gut76}. 
This suggest a weak dependence of this parameter with the projectile mass and energy.

To conclude, compared to the original exciton model existing in the GNASH code, the
approach proposed by Ribansk\'y and Oblo\v zinsk\'y's and implemented in the PREEQ code 
improves greatly the predictions concerning  
complex particle production rates in pre-equilibrium processes, with the adjustement of one free
parameter depending only on the target mass.

\subsection{Exciton model and direct reactions}
In order to modify the original exciton model concerning the complex particle emission
in nucleon-induced reactions, a completely different approach has been proposed
by Kalbach \cite{Kal77}. It is based on the fact that direct reactions such as the 
nucleon pick-up process and the cluster knock-out
process are not included inside the exciton model. Therefore this approach calculates
their associated  contributions separately and add them to the pre-equilibrium component
calculated with the original exciton model. Contrarly to the PREEQ program, this approach 
does not use any multiplication factor in the particle production rate expression and thus, 
it has no adjustable parameter. 
In other words, this approach proposes to replace the cluster formation probability
introduced in Ref. \cite{Ri73} by the contribution of direct reactions. This modification is taken
into account in the code PRECO-2000 \cite{Kal01} which calculates nucleon and complex 
particle non-equilibrium spectra in nucleon-induced reactions using: i) the two-component 
version of the exciton model and ii) phenomenological models for direct reaction processes.
This code is open to the community via the Data Bank Computer Program Services of NEA.
The same approach has been recently implemented in the TALYS code \cite{Kon03},
which is still under development and therefore not yet available to the community.

Calculations have been done with the PRECO-2000 code using the set of global
parameters recommended by the author for the contribution of direct processes.
Details can be found in Ref. \cite{Kal01}. 
For the exciton model
contribution, the same values for specific parameter as for the PREEQ calculations have been used. 
In Fig. 25 an example of the PRECO-2000 results obtained for the alpha-particle
emission in $^{56}$Fe(n,X) reactions at 96 MeV is given. The three individual contributions 
in the non-equilibrium spectrum are displayed. We observe that the very low contribution
of the pre-equilibrium processes predicted by the exciton model (dash-dotted line) 
is compensated by the other two direct processes now included, i.e, the
pick-up of three nucleons (dashed line) and the knock-out of alpha particles (dotted line)
which are assumed to be pre-formed inside the nucleus. The total non-equilibrium spectrum 
is obtained by summing all these contributions.

\begin{figure}
\begin{center}
\includegraphics[width=0.3\textwidth]{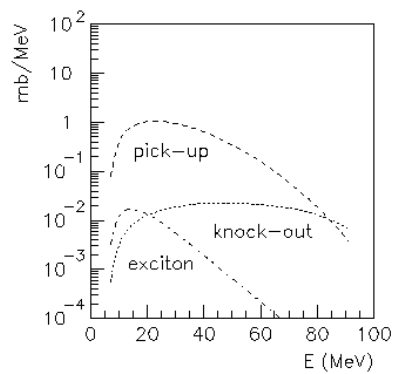}
\caption{\label{fig:epsart}Different mechanism contributions in the non-equilibrium alpha
particle spectrum calculated using the PRECO-2000 code for the $^{56}$Fe(n,X) reaction at 96 MeV.}
\end{center}
\end{figure}

Following the same procedure as in the previous subsection, calculations have been
performed first for the data that we measured at 96 MeV. The results are
presented in Fig. 21 for the $^{56}$Fe(n,Xlcp) and $^{208}$Pb(n,Xlcp) reactions (dashed lines). 
The disagreement with the experimental distributions is rather strong for 
both systems. For the iron case, the non-equilibrium complex particle production
is overestimated while the proton emission is underestimated.
For the lead target, composite ejectile rates are underestimated, as well as the proton
distribution. In addition, for a given target, 
the disagreement seems to become more important as the mass of the emitted particle increases.
Even if the model in PRECO-2000 code predicts more particles in the pre-equilibrium region than GNASH 
does, experimental shapes and amplitudes are not as well reproduced as with the PREEQ code. 
In the case of nucleon ejectiles, the secondary pre-equilibrium emission can be considered 
in this code. However, this contribution was not included in the calculated spectra in
order to get the same calculation conditions as in the 
previous subsection. This can explain the underestimation found for energies around 20 MeV 
in the proton spectra.

\begin{figure}
\begin{center}
\includegraphics[width=0.5\textwidth]{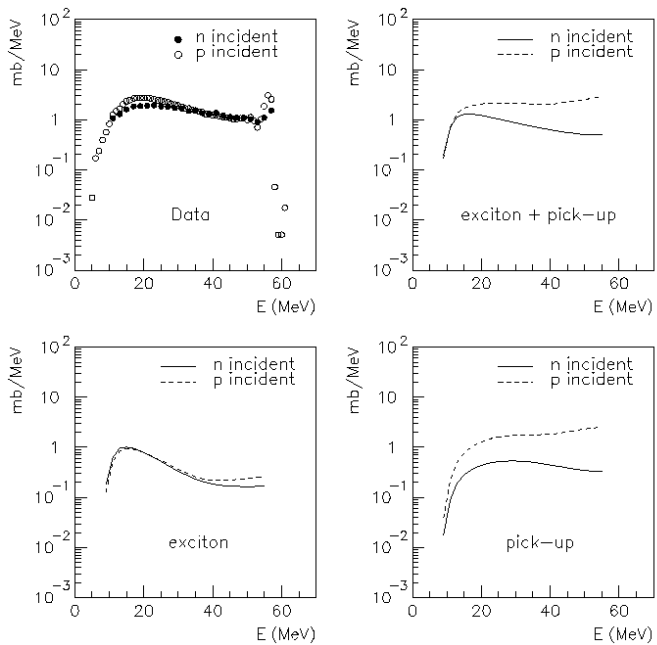}
\caption{\label{fig:epsart}Deuteron emission in proton and neutron induced reactions on 
$^{208}$Pb at 63 MeV. Experimental results (top left panel) are compared to
the distributions calculated using PRECO-2000 code (top right panel). 
Contributions from pre-equilibrium (exciton model) (left bottom panel) 
and nucleon pick-up reaction (bottom right panel) are presented.}
\end{center}
\end{figure}

Despite its bad data reproduction observed at 96 MeV, we tested
PRECO-2000 again by changing the incident particle and energy of the entrance channel.
Doing so, we found a better agreement as it can be seen in Fig. 22, where
the predictions of the PRECO-2000 code (dashed lines) for the 39 MeV $^{209}$Bi(p,Xlcp) 
(top panels) and the 63 MeV $^{208}$Pb(p,Xlcp) (bottom panels) reactions are compared to
the experimental results from the Refs. \cite{Ber73} and \cite{Gue04} (points).
Even if a tremendous disagreement still exists at low incident energies,
the model predictions are sensibly improved with proton projectiles compared to those related
to incident neutrons at 96 MeV. This suggest that the PRECO-2000 predictions strongly depend
on the incident energy and the projectile type. That
latter aspect can be studied in more detail since data with both neutron and proton 
projectiles are available for $^{208}$Pb at the same incident energy (63 MeV).
In Fig. 26, are presented the experimental energy distributions
of deuterons for both reaction types (top left panel): i) (p,xd) \cite{Gue04} (open circles), and
ii) (n,xd) \cite{Ker02} (full circles). The experimental results are very similar in shape and 
in amplitude for both projectiles.
The corresponding PRECO-2000 calculated distributions are shown in the
top right panel. As it can be seen, the theoretical distributions are very different when changing
the incident nucleon type (neutron or proton), in a strong contradiction with the data. 
This disagreement does not originate from the pre-equilibrium contributions calculated by the
exciton model because we checked that the corresponding distributions are similar for 
neutron and proton induced reactions (bottom left panel). On the other hand, the 
calculated contributions for the nucleon pick-up process (right bottom panel) are very different 
from each other and, since this mechanism is dominant, this difference generates the disagreement 
observed with the data. To conclude, the contribution of direct reactions like it is
calculated in PRECO-2000 strongly depends on the incident particle type, contrary 
to the experimental data. This effect constitutes of course a shortcoming of the model.

\subsection{Particle emission at equilibrium}
Compared to PRECO-2000 simulations, the calculations performed with the code PREEQ have shown
that this last approach allows a better description of the particle emission in the 
pre-equilibrium stage. For that reason, the results obtained with this model will be used
in the further discussion. 

As already commented in a previous section, the results presented in
Figs. 21 and 22 suggest that for heavy targets, almost all particles are emitted during 
the pre-equilibrium phase of the reaction. Except for low-energy alpha particles, the PREEQ
calculated distributions allow a good description of the experimental results over the full
energy range, showing that the contribution of the evaporation process should be small. 
On the other hand, for light target nuclei (Figs. 21 and 23), the low-energy component of the
experimental distributions suggest that the particle emission at equilibrium is rather 
important. 

\begin{figure}
\begin{center}
\includegraphics[width=0.5\textwidth]{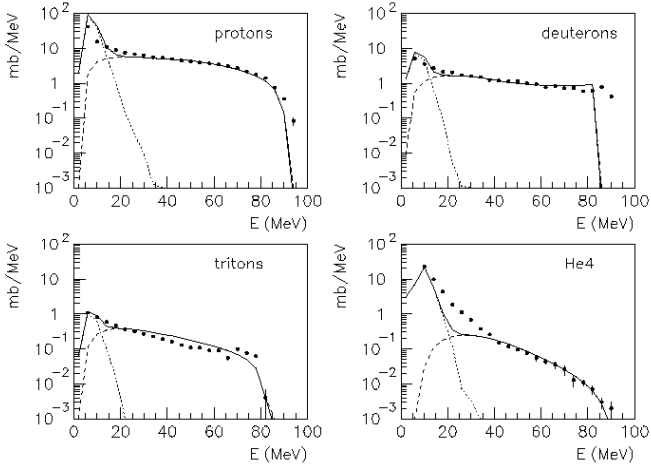}
\caption{\label{fig:epsart}Calculated pre-equilibrium and evaporation contributions 
(dashed and dotted lines respectively) in the particle emission
spectra for the 96 MeV $^{56}$Fe(n,Xlcp) reaction, compared to the experimental results of the present work 
(full circles). The calculated total distributions (sum of pre-equilibrium and evaporation spectrum) 
are presented as continuous lines.}
\end{center}
\end{figure}

In this subsection, we propose to determine the contribution of the evaporation process.
This component can be calculated separately assuming that it results from two 
different sources. The first source is the 
so-called "pure evaporation" and concerns the evaporation from the compound nucleus which has
reached a statistical equilibrium. In Ref. \cite{Wu77}, its contribution is given by a fraction 
f$_{EQ}(E)$=(1-f$_{PE}(E)$) of the total reaction cross section, where f$_{PE}(E)$ is the 
fraction of the pre-equilibrium emission, considering n, p, d, t, $^3$He and alpha particles,
and $E$ is the composite nucleus excitation energy. We determined this fraction using the 
pre-equilibrium spectra calculated with the PREEQ code for all ejectile types. The resulting value
obtained for the 96 MeV $^{56}$Fe(n,X) reactions is f$_{PE}(E)$=0.993, in agreement with that 
estimated for 62 MeV $^{54}$Fe(p,X) reactions in
Ref. \cite{Wu76}. That value very close to 1 shows that almost the entire reaction cross section 
is available for the pre-equilibrium emission, and that the evaporation process of a compound
nucleus represents a very small component with an associated value of f$_{EQ}(E)$=0.007. 
The second source of the equilibrium component which can be considered is the evaporation from
a residual nucleus left in an excited state just after the pre-equilibrium emission has occured. 
In order to estimate  the excitation energy of such a nucleus and its formation probability after the 
pre-equilibrium emission of each outgoing particle type, again, we used the energy
differential distributions previously calculated with PREEQ. The residual nucleus excitation
energy is given by the formula 
$U$=$E$-$B$$_{\beta}$-$e$$_{\beta}$, where $B$$_{\beta}$ and $e$$_{\beta}$ are the binding
energy of the emitted particle ${\beta}$ and its emission energy respectively and $E$ is the 
excitation energy of the compound nucleus.
Having determined that quantity, the evaporation spectra are further calculated using the 
Hauser-Feshbach formalism \cite{Ha52}. Particles are emitted until the evaporation process is no 
longer energetically possible and the nucleus remaining energy is released in the form of gamma rays.

The results obtained for the 96 MeV $^{56}$Fe(n,Xlcp) reactions
are given in Fig. 27 (dotted lines), together with the the pre-equilibrium component calculated with 
PREEQ as described in subsection A (dashed lines). The total particle emission spectrum determined by 
summing both mechanism contributions (continuous line) is also presented and compared to the experimental
data (points). The agreement found over the full energy range is relatively good, except for 
helium isotopes around 20 MeV, where the calculated distributions are below
the experimental results. The same effect has been found for the $^{208}$Pb(n,X$^4$He) reaction, 
showing that the pre-equilibrium
contribution for helium isotopes is underestimated in this energy region for both light and heavy targets. 
For hydrogen isotopes the introduction of the evaporation contribution allows a good description of 
the particle emission over a wide energy range.

\subsection{Angular distributions}
To complete our analysis about the models, we would like to compare the experimental angular 
differential cross sections to the theoritical ones. While the exciton model is largely used
to calculate angle integrated energy spectra, the determination of angular distributions is
out of its capabilities. In order to overcome this difficulty,
several approaches involving modifications of the exciton model have been proposed, like
in Ref. \cite{Ma75}. However, most of them contain serious approximations or induce 
computational complexities and they can be applied only for a limited set of reaction 
configurations.
For this reason, a phenomenological approach proposed in Ref. \cite{Kal88} 
is often preferred to study the continuum angular distributions. It is based 
on a systematical study of a wide variety of experimental data. The parameterization 
established for the double-differential cross section as a function of the total 
energy-differential cross section is given by the equation:
\begin{equation}
\frac{d^{2}\sigma}{d\Omega de}=\frac{1}{4\pi} \frac{d\sigma}{de}
\frac{a}{sinh(a)}[cosh(a cos\theta)+f_{PE}sinh(a cos\theta)]
\end{equation}

In this expression, $\theta$ is the emission angle in the center of mass frame, and
the term $a$ is the slope parameter depending on the incident particle type and energy, 
the target nucleus and the exit channel. It can be calculated using the procedure described 
in Ref. \cite{Kal88}.
The f$_{PE}$ parameter is the fraction of particle emission apart from equilibrium. 
It will be called further the fraction of pre-equilibrium emission and it is
calculated using the formula:
\begin{equation}
\ f_{PE}=\frac{(d\sigma/de)_{PE}}{(d\sigma/de)}
=\frac{(d\sigma/de)_{PE}}{(d\sigma/de)_{PE}+(d\sigma/de)_{EQ}}
\end{equation}
where the $PE$ and $EQ$ symbols refer respectively to pre-equilibrium and equilibrium emissions.
Using the energy-differential cross sections for these two processes calculated in 
subsections A and C, the double-differential cross sections are calculated according to equation (3). 
In Fig. 28 are presented the resulting angular distributions (lines) obtained for the proton
emission in $^{56}$Fe(n,X) and $^{208}$Pb(n,X) reactions at 96 MeV (right and left figures respectively),
together with the experimental data (points). In order to have a better illustration of 
the different reaction mechanisms 
which contribute to the particle emission spectra (evaporation and pre-equilibrium emission), 
when we constructed the angular distributions, we chose three different energy domains: 
$8-12$ MeV (continuous lines), $40-44$ MeV (dashed lines) and $68-72$ MeV (dotted lines). 
The contribution of the pre-equilibrium emission in the total cross section (f$_{PE}$ factor)
for each domain is also given near the corresponding distribution.

\begin{figure}
\begin{center}
\includegraphics[width=0.5\textwidth]{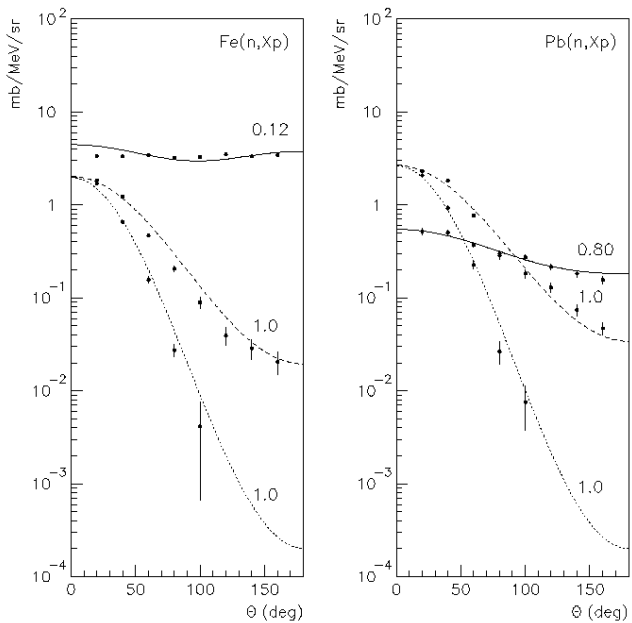}
\caption{\label{fig:epsart}Double-differential distributions calculated using the
parameterization from Ref. \cite{Kal88} (lines) for proton emission in 96 MeV 
neutron-induced reactions on $^{56}$Fe and $^{208}$Pb compared with the experimental results (points).
The continuous, dashed and dotted lines correspond to 
the $8-12$ MeV, $40-44$ MeV and $68-72$ MeV emission energy ranges respectively.
The contribution of the pre-equilibrium emission in the total cross section (f$_{PE}$ factor) for each 
domain is also given near the corresponding distribution.}
\end{center}
\end{figure}

We observe in general satisfactory agreement between the theoretical results and the 
experimental distributions. At low energies ($8-12$ MeV), particles are emitted from both
evaporation and pre-equilibrium processes whose respective contributions depend
on the target nucleus mass. For the iron case, we found f$_{PE}$=0.12 
and we observe a quasi-isotropic distribution, both signals indicating that the evaporation 
process is dominant for light targets. For the lead target, f$_{PE}$=0.80 and the angular 
distribution is slightly forward peaked, showing that low energy particles
are mainly emitted during the pre-equilibrium stage. 
For more energetic particles, f$_{PE}$=1 for both targets, and we observe that
they are mainly emitted at small angles, following the beam direction. From this, we 
deduce that those ejectiles are emitted before an equilibrium has been reached.
We found a similar agreement when we built the complex particle distributions, showing 
that the Kalbach parameterization is able to give a proper description of the double 
differential cross sections, whatever the target or the emitted particle are. In addition, a 
physical basis for this parameterization has been established
in Ref. \cite{Ch94}, allowing a more detailed theoretical understanding of the properties 
of the angular distributions in the continuum energy domain.

\section{Summary}
In this paper, we report a new set of experimental data concerning light charged-particle production in
96 MeV neutron-induced reactions on natural iron, lead and uranium targets. Double-differential 
cross sections of charged particles have been measured over a wide angular range 
($20-160$ degrees). With the MEDLEY set-up, data were measured for p, d, t, $^3$He and alpha 
particles, with low energy thresholds. The SCANDAL set-up has been
used to measure proton production cross sections in the same angular range, with good statistics and 
angular resolution, but with an energy threshold of about 30 MeV. For proton emission, the very
good agreement found between both sets of measurements obtained with both independent detection
systems shows that we had a good control on the systematical uncertainties involved.
This is due, in part, to the unambiguous cross section normalization which has been applied
using very accurate data on the \textit{np} scattering cross section \cite{Ra01}. 
In our experiment, we also measured this cross section and we obtained a good agreement
with data from Ref. \cite{Ra01}. The estimated systematical uncertainties affecting the double differential
cross sections reported in this work are of the order of 5 \%.

Data presented in this paper allows the extension to higher energies (up to $96$ MeV) of the available
experimental results on nucleon-induced reactions in the $20-200$ MeV energy range, which were up to now 
limited to about 60 MeV incident energy.
This new data set, together with the data already existing in the literature, allows us to study in 
detail both main theoretical approaches \cite{Ri73,Kal77} available nowadays for the 
description of nucleon and complex particle emission in nucleon-induced reactions at intermediate
energies. 
These approaches have been proposed mainly to improve the exciton model predictions concerning 
the production of clusters, which was originally strongly underestimated by the model, as shown with 
the calculations we have performed with the GNASH code \cite{Ch92}. 
Since the cross sections evaluated with GNASH 
are at present implemented in the MCNPX code, we would like to issue a warning that some important 
information needed in specific application as the power deposited in a spallation target of an ADS 
could be underestimated.

In order to test both approaches, we performed calculations with the PREEQ \cite{Bet75} 
and PRECO-2000 \cite{Kal01}
codes. The PREEQ results have shown that by taking into account the cluster formation probability 
in the pre-equilibrium stage of the reaction, one can obtain a global agreement over a wide set of
configurations. The formation probability is a free parameter in the PREEQ code and we 
have adjusted it for each target nucleus. The evolution of the resulting values shows that, for a 
given outgoing particle, the probability decreases with the target mass. In addition, 
for a given target, the formation probability is larger for lighter
particles. This parameter depends very little on the incident particle type
and energy. Proposed as an alternative to this approach, the method used in the PRECO-2000 code and
implemented in the more recent code TALYS \cite{Kon03} to calculate
complex particle production cross sections, considers contribution of direct reactions in the outgoing
spectra. In many cases, however, this approach does not lead to a good reproduction of the experimental 
distributions. Despite the acceptable agreement found in some particular situations, it can not be used
for the moment in a global description of nucleon-induced reactions. This deficiency is due in part to 
the strong dependence of its predictions on the projectile type. It is our hope that the work performed
at present in the development of the TALYS code will soon provide an improved version of this 
approach.

We have completed the description of the particle emission over the full energy range by adding the 
contribution of the evaporation process to the pre-equilibrium emission calculated using the PREEQ code.
That calculation scheme has shown that for heavy targets, almost all particles are emitted during 
the pre-equilibrium stage of the reaction, while for light targets, a strong component from the 
evaporation process is present at low emission energy. In
addition, the most important contribution in the equilibrium component originates from the decay of
residual nuclei left in an excited state after the pre-equilibrium particle emission.
Finally, we shown that a correct description of the energy-differential distributions and of the 
different mechanisms contributing to the total cross section allows to calculate double-differential 
cross sections by including also the parameterization from Ref. \cite{Kal88} for the angular
distribution determination.
The good reproduction of the shapes of the double-differential distributions that we
obtained with this method, suggests that theoretical models must provide at least a good
description of the energy-differential cross sections. The parameterization established 
in Ref. \cite{Kal88} allows a more detailed study of the reaction with a rather satisfactory accuracy 
by allowing the prediction of the double differential distributions.

The results presented in this work show that the understanding of nucleon-induced reactions at these 
energies is far from complete. Two approaches are available in the framework of the exciton model 
for the description of cluster emission in these specific reactions and among them, only that
based on the coalescence model seems to have a satisfactory predictive power. It is, however, based on a
scale factor associated to the formation probability of complex particles which has to be adjusted to
experimental data. Therefore, further theoretical progress 
must be done in this field in order to improve the existing theoretical approaches of the exciton model
and to provide new models based on different considerations. An alternative has been recently 
proposed in this direction, which uses the wavelet technique to simulate the nuclear dynamics and whose
results are very encouraging. They will constitute the subject of a future publication \cite{Se04}.   

\begin{acknowledgments}
This work was supported by the European Community under the HINDAS project 
(contract No. FIKW-CT-2000-0031), the GDR GEDEON (research group CEA-CNRS-EDF-FRAMATOME), 
Vattenfall AB, the Swedish Nuclear Fuel and Waste Management Company, the Swedish Nuclear Power Inspectorate, 
Barseb{\"a}ck Power AB, Ringhals AB, the Swedish Defence Research Agency and the Swedish Research Council. 
We would like to thank the TSL staff for assistance and quality of the neutron beam. 
We are also grateful to Dr. E. Betak for very useful discussions concerning calculations with the PREEQ code. 
Special thanks to Dr. C. Kalbach for her thirty years of effort which have contribute significantly to the 
progress of theory in nucleon-induced reactions.
\end{acknowledgments}

\end{document}